\documentclass[article,onecolumn,12pt]{IEEEtran}

\usepackage{hyperref}
\usepackage{float}[H]
\usepackage[top=3cm, bottom=2.5cm, left=2.5cm, right=2.5cm, headheight=15pt]{geometry}
\usepackage[pdftex]{graphicx}
\usepackage[caption=false,font=footnotesize]{subfig}
\usepackage{amsmath}
\usepackage{amssymb}
\usepackage{amsthm}
\usepackage{amsfonts}
\usepackage{breqn}
\usepackage{lipsum}
\usepackage[english]{babel}
\usepackage[autostyle, english = american]{csquotes}
\usepackage[bibstyle=authoryear, citestyle=authoryear, backend=bibtex, maxbibnames=99]{biblatex}
\MakeOuterQuote{"}
\usepackage{tikz}
\usetikzlibrary{trees}
\usetikzlibrary{arrows.meta}

\bibliography{Bib.bib}
\providecommand{\keywords}[1]
{
  \small	
  \textbf{\textit{Keywords ---}} #1
  \normalsize
}

\usepackage{fancyhdr}
\begin{document}
\bstctlcite{IEEEexample:BSTcontrol}


\title{\textbf{\Large Basically Trapped}\\
\Large\emph{A General Equilibrium Approach to the Undersupply of Basic Research in Free Markets}\\[0.7cm]
{\normalsize Working Paper}
}

\author{
        \normalsize{Lucas D. KONRAD}\\[0.5cm]
        \textit{Vienna University of Economics and Business (WU)}\\
        \textit{Department of Economics}\\
        \textit{Welthandelsplatz 1, 1020 Vienna, Austria}\\
        \textit{Email: lucas.konrad@wu.ac.at}
}

\maketitle
\vspace{-2em}
\noindent\rule[0.5ex]{\linewidth}{0.5pt}
\begin{center}
\textit{\small This is a preliminary version. Please do not cite without permission.\\
Comments and suggestions are welcome.}
\end{center}
\noindent\rule[0.5ex]{\linewidth}{0.5pt}

\vspace{1em}

\begin{abstract}
\noindent In this paper I propose a micro-based innovation driven general equilibrium growth-model allowing for endogenous entry and exit as well as three different types of research. I make the novel distinction between three types of firms, namely \textit{basic} and \textit{applied} research where applied research is again differentiated into high and low type research to account for a heterogeneous firm environment. I further propose a new and flexible way to model basic research spillovers in this model class. While previous literature addresses the effect of R\&D policies in general, my findings suggest, there is no optimal market solution in which basic research takes a noteworthy share. In order to counteract this undersupply and to take advantage of basic research spillovers, a sizable basic research subsidy is required to sufficiently crowd-in basic research which then allows for a substantial increase in welfare. My findings also suggest the necessity for purely state-financed basic research to approach the social planner solution in a feasible manner.
\end{abstract}

\keywords{Industrial Economics, General Equilibrium, Basic Research, Creative Destruction}

\vspace{1cm}

\noindent\rule[0.5ex]{\linewidth}{1pt}
\vspace{-1em}
\renewcommand{\abstractname}{}
\begin{abstract}
\noindent\textit{Acknowledgements:} I thank the Dutch Bureau of Economic Policy Analysis for facilitating my project and in particular and most sincerely Yannis Kerkemezos who accompanied this project in detail and provided valuable critique and support. Further, I express my gratitude to Bastiaan Overvest with whom I developed the thematic outline of the project, as well as my university supervisor Christoph Walsh who contributed valuable feedback throughout the progress of this paper. The views expressed in this paper are those of the author and do not necessarily reflect those of Vienna University of Economics and Business. All errors are my own.
\end{abstract}

\newpage
\setcounter{page}{1}

\section{Introduction}\label{sec:Int}

The size and type of industrial policies are one of the key responsibilities of policy makers. As innovation is a well recognized driver of firm performance and economic growth in general, a central problem lies in finding policies that yield a cost-efficient solution to improve research conducted by private firms. Policy makers to the present show a tendency to subsidize incumbent firms. Evidence suggests that for example most pan-European subsidies are granted to large corporations due to claimed efficiency advantages (\cite{criscuolo2012causal}). The efficiency of such policies is highly debated as on the one hand they might encourage the investment undertaken by incumbents and foster competition and thus lower entry barriers which leads to an increase in overall productivity (\cite{aghion2015industrial}). On the other hand they might harm the reallocation of capital and labor which is considered to be an important driver of economic growth (\cite{haltiwanger2001aggregate}). 

Modelling and estimating effects of redistributional policy interventions have only recently started to attract the attention of scholars. The heterogeneity of firms has to be recognized to accurately capture the effects of such measures. The distinction between firms with a high innovation capacity (high type) and firms with a low innovation capacity (low type) was first made by \textcite{acemoglu2018innovation}. The authors establish a connection between young and highly innovative enterprises as well as old and less innovative ones. Similarly, the distinction between basic and applied research is suggested in the literature (e.g. \cite{akcigit2021back}). Whereas basic research is not directed towards a certain goal, applied research has a clear intention for the application of research. While basic research is commonly supposed to be supplied by the government, the share of basic research conducted by the private sector is sizable and estimated to be around 22\% in the US (\cite{howitt2000economics}) which makes it a substantial characteristic of the private research landscape.

I combine the two above mentioned distinctions and develop a model allowing for three types of firms, namely firms conducting basic research, highly innovative applied firms and less innovative applied firms to allow for a more heterogeneous firm environment and to gain insights over the contribution of each firm type to economic growth in general and the effect of R\&D subsidies on those types in particular. I thereby attempt to contribute to the understanding of the role of different research types in a market economy and gain insights over the welfare effects of research subsidies in such a setup.

In particular, I adopt the general equilibrium model of \textcite{acemoglu2018innovation} by firms doing basic research, and thus alter the two firm type economy they consider to a three firm type economy. This additional dimension addresses a shortcoming of \textcite{acemoglu2018innovation} as a distinction between highly innovative (young) firms and relatively uninnovative (old) firms is of questionable sufficiency considering the above mentioned research environment in the private sector.


I model two distinct key characteristics of basic research, following \textcite{akcigit2021back}. The first is that basic research has larger spillovers, thus effecting the productivity of R\&D by applied firms. This follows the reasoning that an innovation which is not focused on the completion of a specific goal has effects overarching industries. If a company manages to innovate doing basic research, many firms can take advantage of this innovation, regardless of the sector the applied research firm operates in. Such a 'game-changer' allows for related applied research which increases their productivity. I propose a novel way to create a dependency between the amount of basic research in an economy and the size of the spillovers. Larger spillovers together with the quasi-public good nature of information contributes to the often discussed shortage in supply of basic research, as recognized by economic research institutions such as e.g. the Joint Economic Committee (\citeyear{jec2019report}).
Secondly, as costs of innovation cannot be incorporated by it's founder this leads to lower expected return on basic research innovation. Consequently, given the same research intensity, basic research is more expensive. Adding to this more costly nature of basic research is the relatively high level of uncertainty with respect to the research outcome (e.g. \cite{rosenberg2010firms}).

The modelled economy is supplied by firms who innovate on goods to increase productivity. Firms can employ three different types of research, namely basic research, and two types (high and low innovative) of applied research to account for heterogeneity in applied product innovation. The type of research conducted for a given good is determined at the stage of entry. The intuition here is that some goods are better suited for a given type of research than others.

The model further makes a distinction between skilled and unskilled labor. Unskilled labor is concerned with the production of a given good, which is insensitive to the type of research conducted on it. Skilled labor serves two purposes. First, a given share of skilled labor is required to enable the production process, which can be considered as labor employed in management or accounting. This amount of managing labor required is the same for all product lines, which is in line with the identical production of goods and is part of the fixed costs of a firm. Secondly, skilled labor is the driver of R\&D. A firm decides to employ a given amount of skilled labor to improve the productivity of goods outside their production repertoire. This innovation, if successful, results in the acquisition of that product. The decision over the amount of skilled labor required depends on the type of research employed by that firm.

After successful firm entrance, highly innovative applied firms over time lose some of their innovation capacity. The resaoning is here that even if a good is highly improvable by applied research, this improvability is eventually exploited to an extend where profitability gains can only be gained at a higher cost, or at a lower rate. This effect is discussed by \textcite{acemoglu2018innovation} who establish a connection between highly productive R\&D and young (small) firms as well as poorly productive R\&D and large (old) firms.

Basic research, similar to highly innovative applied research, can also transition to a low applied research type, albeit with a different interpretation. Basic research, once successfully performed, has the ability to become \textit{outdated}. As a basic innovation happens, applied firms can benefit from this through spillovers. However, after some time, the extra innovation capacity is exhausted as the basic innovation becomes public knowledge. At this stage, previously basic research firms transform into low applied research firms. 

The key interest of this study is to analyze the equilibrium and welfare gains of policies once the model of \textcite{acemoglu2018innovation} allows for basic research. Analyzing the market equilibrium as well as the effect of R\&D subsidies for specific research types allows to analyze the potential of research-type-specific subsidies and solving for the social planner optimum provides an insight into the potential of such measures. My findings suggest that the market solution undersupplies high-applied research, and basic research even more so. The social planner can overcome this issue for high-applied but not for basic research by imposing an environment which encourages redistribution of labor and encourages exit of low-applied research. For basic research and thus for the associated spillovers to take effect, the economy first requires substantial incentives from the policy maker to attract basic research firms by means of a basic research subsidy (crowding-in). Once such a policy is in place, the equilibrium growth path is considerably altered due to the overarching positive effect on all types of research. The social planner allows for an even greater welfare benefit if basic R\&D policy for incumbents is in place. Optimizing the innovation environment accordingly where high-applied and basic research are almost unconstrained in their operation and low-applied firms are encouraged to exit to assure efficient redistribution. As the social planner imposes almost no entry barriers for basic research firms, my findings also suggest the necessity of fully governmental financed research to achieve this optimum in a feasible manner.

My model constitutes an amendment of \textcite{acemoglu2018innovation} inspired by \textcite{akcigit2021back}.
\textcite{acemoglu2018innovation} apply a general equilibrium model based on US firm data. The authors find that an incumbent subsidy provides a higher welfare increase than an entry subsidy of the same size. However, the social planner would take advantage of redistributing labor from less efficient firms to more efficient ones, substantially outperforming the welfare increase of an incumbent subsidy.
\textcite{akcigit2021back} analyze the effect of basic research in a general equilibrium model on economic growth calibrated on French data from 2000 to 2006. The authors find a tendency of general R\&D subsidies in the private sector to relatively over-subsidize applied research, leading to a miss-allocation in skilled labor. However, subsidizing government funded basic research improves welfare substantially.
Both of the above mentioned papers are based on the the technology driven endogenous growth literature, starting with \textcite{romer1990endogenous}, \textcite{aghion1990model}, \textcite{grossman1991quality} and in particular \textcite{klette2004innovating}, \textcite{lentz2008empirical} who provide the model structure then used by \textcite{acemoglu2018innovation} and \textcite{akcigit2021back} to who this work relates to the closest.

The subsequent paper is structured as follows: I will explain the model in section \ref{sec:Mod}, then I will discuss the computational algorithm to solve the mode in section \ref{sec:Comp}. In section \ref{sec:Res} and \ref{sec:Dis} the model implied results are discussed and possible caveats and shortcomings are outlined. Finally, section \ref{sec:Con} concludes.

\section{The Model}\label{sec:Mod}

In the following section I will discuss the model which leads to an equilibrium growth path, the imposed characteristics on the economic actors and the production process, as well was the model dynamics allowed for and the optimal decision rules associated with them. The model is closely related to \textcite{acemoglu2018innovation} which is why for clarity the notation is adopted where possible. The essential amendments are explicitly mentioned. 
Note for clarity that the subsequently modeled economy is closed with a continuum of goods being produced out of the set $[0,1]$. This further implies a continuum of firms, only a subset of which are producing at each point in continuous time.

\subsection{Economic Actors}

The representative household is assumed to have an infinite lifespan and a constant relative risk aversion (CRRA) utility function 
\begin{equation}
    U_0=\int_{0}^{\infty}{e^{-\rho t}\frac{C\left(t\right)^{1-\vartheta}-1}{1-\vartheta}dt}
    \label{eq:utility}
\end{equation}
where $C(t)$ constitutes the aggregate consumption taking the form 
\begin{equation}
    C\left(t\right)=\left(\int_{\mathcal{N}\left(t\right)}{c_j\left(t\right)^{\frac{\varepsilon-1}{\varepsilon}}dj}\right)^{\frac{\varepsilon}{\varepsilon-1}}.
    \label{eq:agg_cons}
\end{equation}
In equation (\ref{eq:utility}) and (\ref{eq:agg_cons}) respectively $\rho>0$ represents the agents discount rate of future consumption and $\vartheta$ is the inverse of the inter-temporal elasticity of substitution. Further, $\mathcal{N}(t)\subset[0,1]$ is the set of active goods in the economy at time $t$, $c_j(t)$ is the consumption of good $j$ and lastly $\varepsilon>1$ constitutes the elasticity of substitution between goods. Note that the consumption aggregate $C(t)$ is the numéraire in the economy.

The supply of labor is considered to be inelastic and is distinguished into 'skilled' and 'unskilled' labor. Unskilled labor is employed in the production process which is unsensitive to the firm type. Skilled labor takes two functions, first each operating firm requires a given amount $\phi>0$ of skilled labor in management ($L^M$) in order to be able to operate the firm. Also, firms employ innovating skilled labor ($L^{R\&D}$) which is concerned with R\&D, i.e. innovation on goods. Consequently, labor demand takes the form
\begin{equation}
    L^P(t)=1 \quad\text{and}\quad L^M(t)+L^{R\&D}(t)=L^S
    \label{eq:labor}
\end{equation}
where $L^P$ and $L^S$ is the amount of unskilled and skilled in the economy labor respectively. Note that the amount of unskilled labor is normalized to one to assure identification.

Households then maximize utility in equation (\ref{eq:utility}) w.r.t. their budget constraint
\begin{equation*}
    \Dot{A}(t)+C(t) \leq r(t)A(t)+L^P w^u(t)+L^S w^s(t)
    \quad\text{s.t.}\quad 
    \int_0^\infty e^{-r(t)t}A(t)dt\geq 0.
\end{equation*}
Here, $A(t)$ states the assets held by the household at time $t$, $\Dot{A}$ is the change in $A(t)$ over time ($\frac{\partial A}{\partial t}$), $r(t)$ is the equilibrium interest rate of the economy and $w^u$ and $w^s$ are the wages for each labor type. Finally the no-Ponzi condition must hold, i.e. the present value of all discounted future payoffs must be non negative.

Finally I assume in line with \textcite{acemoglu2018innovation} that the economy is closed and thus it holds that
\begin{equation}
    Y(t)=C(t).
\label{eq:prod_is_cons}
\end{equation}

The CRRA utility function in (\ref{eq:utility}) further yields the Euler equation
\begin{equation}
    \frac{\Dot{C}}{C}=\frac{r-\rho}{\vartheta}.
    \label{eq:euler_eq}
\end{equation}

\subsection{Good Production}

Each good $j$ is produced by firm $f$ which has the leading technology in that good. The leading technology is obtained by being the firm which most recently successfully innovated on that good. Consequently, firms producing one or more goods at time $t$ are \textit{active} in the economy, and firms producing nothing are \textit{inactive}. I denote the set of active firms $\mathcal{F}$. Note that the model does not distinguish between one firm producing multiple goods and each sub-firm producing one good, as superadditive effects in firm behaviour are excluded by assumption. This leads to the property that a firm holding and producing $k$ goods can be split into $k$ firms producing one good as will be shown below in detail below.

\begin{figure}
    \centering
\normalsize
\tikzstyle{every node}=[draw=black,thick,anchor=west]
\begin{tikzpicture}[%
  grow via three points={one child at (0.5,-0.7) and
  two children at (0.5,-0.7) and (0.5,-1.4)},
  edge from parent path={(\tikzparentnode.south) |- (\tikzchildnode.west)}]
  \node {Exp. Return on Innov. Positive? (Y/N)}
    child { node {(N) No Innovation}}
    child { node {(Y) Innovation in 1 Random Good}
        child { node {Pioneer? (Y/N)}
            child { node {(Y) Introduce Good as Monopolist}}
            child { node {(N) Degenerate Price Competition}
                child { node {Entrant Prevails}}
            }
        }
    };
\end{tikzpicture}
\
\caption{Decision tree for firms: Each firm faces the decision whether or not to innovate at each point in time. If innovation is successful the innovator will either introduce an inactive good to the market or takeover an active good, depending on which good the firms innovates on.}
\label{fig:innov_tree}
\end{figure}
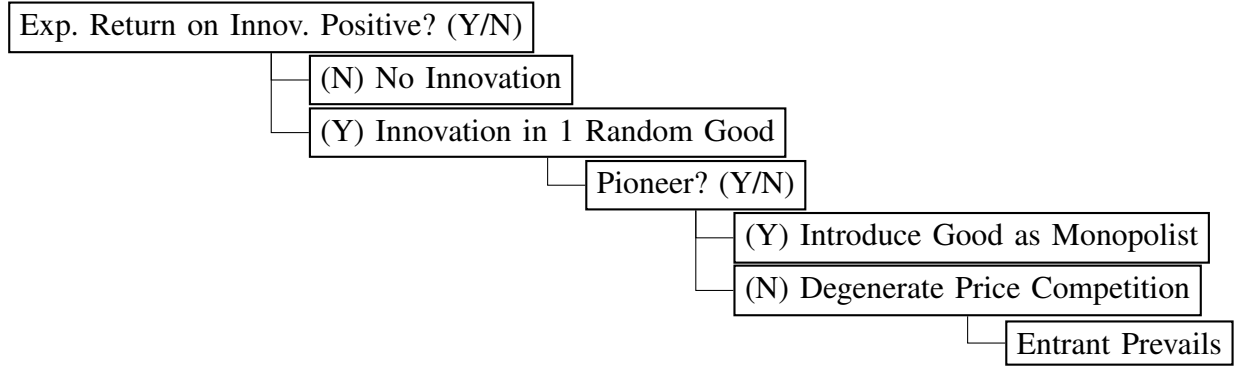

The innovation decision of a firm depends on the expected return on innovation. If Innovation is expected to be profitable a random good among all goods (also inactive ones) is innovated on.\footnote{Note that with splitting firms into sub-firms producing one good, the probability of innovating on the currently produced good is zero.} If this good is profitable after innovation, the good is either introduced into the market or taken over from a competitor if it was already produced. As it is not self evident that the entrant is outperformed, both firms enter a degenerate price competition with no entry costs in which the more productive one prevails, i.e. the recent innovator. Consequently, successfully innovating on an active good always drives the respective incumbent out of the market. This mechanism is illustrated in figure \ref{fig:innov_tree}.

A uniform linear production function is assumed for all firms in the market taking the form
\begin{equation}
    y_{f,j}=q_{f,j}l_{f,j}
    \label{eq:prod_fun}
\end{equation}
where $y_{f,j}$ constitutes the value of the produced good $j$ for producing firm $f$, $q_{f,j}$ is the level of productivity for that good and $l_{f,j}$ is the number of producing (unskilled) labor required for production. Thus, if firm $f$ holds $n$ products, the set of productivities is defined as

\begin{equation}
    \mathcal{Q}_f\equiv\left\{q_{f,j_1},\ldots,q_{f,j_n}\right\}.
    \label{eq:prod_set}
\end{equation}

Note that given equation (\ref{eq:prod_fun}) we obtain the firms' cost function for the production process to be
\begin{equation*}
    \kappa=lw^u=y\frac{w^u}{q}\Rightarrow MC = \frac{\partial k}{\partial y}=\frac{w^u}{q}
\end{equation*}
where MC is the marginal cost of production. As $w^u$ applies for the whole economy and $q$ is non-decreasing in $t$, firms produce by considering relative productivity defined as
\begin{equation}
    \hat{q}\equiv\frac{q}{w^u}=\frac{1}{MC}.
    \label{eq:rel_prod}
\end{equation}

Finally, the productivity of the economy is represented by the Dixit-Stiglitz-aggregator of productivities over all active goods $\mathcal{N}$, taking the form
\begin{equation}
    Q\equiv\left(\int_{\mathcal{N}}{q_j^{\varepsilon-1}dj}\right)^{\frac{1}{\varepsilon-1}}.
    \label{eq:prod_index}
\end{equation}

\subsection{Basic \& Applied Firms}

The model allows for three types of firms operating in the economy, which is novel in the literature. Firms can employ basic or applied research following \textcite{akcigit2021back}. To further model the heterogeneity of firms as outlined by \textcite{acemoglu2018innovation} applied research firms are distinguished into highly innovative firms and low innovative firms. 

As outlined above and illustrated in figure \ref{fig:innov_tree} an entrant randomly selects a good to innovate on. If this good is not already produced (i.e. if it is inactive) the type of the good is unknown and hence the research conducted on it is unknown. An entrant selects its type at the time of entry, depending on the expected properties of the good it innovates on. The probability distribution of selecting a certain type is assumed to be
\begin{equation}
    Pr(\theta = \theta^{a,l}) = (1-\alpha)(1-\beta) \quad Pr(\theta = \theta^{a,h})=(1-\alpha)\beta \quad Pr(\theta = \theta^b)=\alpha
    \label{eq:entry_prob}
\end{equation}
where $\theta^{a,l}$ and $\theta^{a,h}$ correspond to the innovation capacities of low productive and highly productive applied firm types respectively and $\theta^b$ is the innovation capacity of firms conducting basic research. Further, $\alpha$ and $\beta$ are the probabilities of conducting basic research and highly innovative applied research respectively. The innovation capacities as well as the firm type probabilities are exogenously derived. Considering the two applied research types, intuition aligns with the assumption that $\theta^{a,l}<\theta^{a,h}$ which is in line of the finding of \textcite{acemoglu2018innovation}. As outlined above a key property of basic research is that it is expensive and risky to do as payoffs are uncertain and implementation of the new technology is expensive. The resulting risk premium and cost-markup due to implementation reflects in the assumption that $\theta^{b}<\theta^{a,l}$ which is supported by \textcite{akcigit2021back}.  Consequently the inequality $\theta^{b}<\theta^{a,l}<\theta^{a,h}$ is assumed to hold.

\subsection{Innovation}

Innovation is performed by all types of firms on active and inactive goods. However, they vary in innovation size and the types of spillovers at each time point. The rate of successful innovation within a firm takes the structure of
\begin{equation}
    X_f = \theta_f^\gamma n_f^\gamma h_f^{1-\gamma}
    \label{eq:inov_intens}
\end{equation}
in which $X_f$ is the flow rate of adding a new (\textit{active}) product to the economy. Further, $\theta_f$ is the innovation capacity of that firm and $n_f$ is the number of product lines produced at a given point in time. Lastly, $h_f$ is the amount of skilled labor it has to hire to achieve that flow rate $X_f$ and $\gamma\in(0,1)$ constitutes the innovation elasticity with respect to R\&D, or explicitly, by what percentage the innovation rate is increased if a firm increases R\&D expenditure by 1\%. Note that for entrants $n_f=1$. In order to normalize over firm size let $x_f \equiv \frac{X_f}{n_f}$, which can be interpreted as the innovation rate of firm $f$ per good. Below I show that the assumption of $n=1$ can be generalized to all firms which then allows to establish the connection between goods that encourage a certain type of research and a firm conducting this type.

Please not that for clarity, the subsequent notation ignores the firm specific subindex $f$ and the time specific subindex $t$ if it causes no confusion. Each equation holds at each point in time for each firm individually if not stated otherwise. 

With equation (\ref{eq:inov_intens}) at hand, the cost of R\&D for a given firm is
\begin{equation*}
    w^sh = w^sn\left(\frac{h^{1-\gamma}\theta^\gamma n^\gamma}{n}\right)^{\frac{1}{1-\gamma}}\theta^{-\frac{\gamma}{1-\gamma}}=
    w^s n x^{\frac{1}{1-\gamma}}\theta^{-\frac{\gamma}{1-\gamma}}
\end{equation*}
leading to the R\&D cost function
\begin{equation}
    C(x,n,\theta) = w^s n G(x,\theta) \quad \text{for} \quad G(x,\theta)\equiv x^{\frac{1}{1-\gamma}}\theta^{-\frac{\gamma}{1-\gamma}}
    \label{eq:rd_cost}
\end{equation}
where $G(x,\theta)$ takes the interpretation of the amount of skilled labor required in order to achieve an innovation rate of $x$ given a research type $\theta$.

If innovation is successful for a product line (which happens at rate $x$) the productivity of the good is enhanced. Note that firms do not know, on which good they successfully innovate a priori and consequently there expected level of productivity is
\begin{equation*}
    \Bar{q} = \int_0^1 q_j dj
\end{equation*}
Assuming (for now) no spillovers (as in \cite{acemoglu2018innovation}), the expected productivity after innovation on product $j$ is modeled as
\begin{equation}
    q_j(t+) = 
    \begin{cases} 
        q_j(t) + \lambda \Bar{q}(t)\quad &\text{for} \quad \theta\in\{\theta^{a,l},\theta^{a,h}\}\\
        q_j(t) + \eta \Bar{q}(t) \quad &\text{for} \quad \theta = \theta^b
    \end{cases}
    \label{eq:inov_st_no_spill}
\end{equation}
where $\lambda$ and $\eta$ are the innovation step sizes for successful applied and basic research respectively, both of them entering the model exogenously. Note that intuition would suggest $\lambda<\eta$ as successful innovation in basic research can be assumed to have a larger impact on productivity. Also notice the implied market failure that each innovating firm can built on top of the existing technology for a given good. Dealing with continuous time, $t+$ constitutes the time just after t.

However, a known property of basic research is that it has larger spillovers meaning that basic innovation in one good allows applied researchers to freeride on it and increase the profitability of their own research without additional costs. This key feature finds its place in the model by allowing for a dependency between the innovation step size of applied research and the active basic research in the economy. This assumption meets intuition as more basic research creates more spillovers for more applied firms. If however goods are currently inactive on which basic research would be conducted on, those do not contribute to spillovers. With equation (\ref{eq:inov_st_no_spill}) at hand, I propose to model this dependence of applied innovation steps on basic research linearly\footnote{Note that the linearity assumption is not backed by theory or empirical observations. Due to the lack of data for this project it constitutes the most simple solution, however, the structural assumption on the interdependence between innovation step sizes can take any structure, e.g. convex in basic research share, allowing for interaction effects between basic research achievements. If this project is developed further, the structure of this spillovers will be a central element of interest.} as the post spillover innovation step size is a weighted average between the (high) basic research step size $\eta$, and the pre-spillover step size $\lambda$, taking the structure of
\begin{equation}
    q_j(t+) = 
    \begin{cases} 
        q_j(t) + (\xi\eta + (1-\xi)\lambda) \Bar{q}(t) &\text{for} \quad \theta\in\{\theta^{a,l},\theta^{a,h}\}\\
        q_j(t) + \eta \Bar{q}(t) &\text{for} \quad \theta = \theta^b
    \end{cases}, \text{for}\quad \xi = \frac{\Phi^b}{\Phi^{a,l}+\Phi^{a,h}+\Phi^b}.
    \label{eq:inov_st_with_spill}
\end{equation}
Here, $\Phi^b$ is the share of active firms in the economy conducting basic research and $\Phi^{a,l}$+$\Phi^{a,h}$ is the share of active firms conducting applied research. Thus, $\xi$ constitutes the relative share of basic research among active firms. If no basic research is applied ($\xi=0$) at a given time $t$, the innovation behaviour breaks down to equation (\ref{eq:inov_st_no_spill}).

\subsection{Prices \& Profits}

The inverse demand function takes the shape of
\begin{equation}
    p_j = C^{\frac{1}{\varepsilon}} c_j^{-\frac{1}{\varepsilon}}
    \label{eq:price}
\end{equation}
And given the observation from equation (\ref{eq:rel_prod}) together with the price function in equation (\ref{eq:price}) yields the profit maximization for each product line
\begin{equation}
    \pi(\hat{q}_j) = \max_{c_j\leq 0}{\left\{\left(p_j-MC_j\right)c_j\right\}} = \max_{c_j\leq 0}{\left\{\left(C^{\frac{1}{\varepsilon}} c_j^{-\frac{1}{\varepsilon}}-\hat{q}^{-1}_j\right)c_j\right\}}
    \label{eq:profit}
\end{equation}
which when maximized provides the optimal consumption $c_j^*$ and optimal price $p_j^*$ by finding
\begin{equation}
    p^*_j = \frac{\varepsilon}{\varepsilon-1}\hat{q}^{-1}_j \quad \text{and} \quad c_j^*=\left(\frac{\varepsilon-1}{\varepsilon}\right)^\varepsilon C \hat{q}_j^\varepsilon.
    \label{eq:eqi_price_cons}
\end{equation}
Using the above result in equation (\ref{eq:profit}) and (\ref{eq:eqi_price_cons}), equilibrium profits are
\begin{equation*}
    \pi^*(\hat{q}_j) = \frac{\varepsilon}{\varepsilon-1}\left(\frac{\varepsilon-1}{\varepsilon}\right)^\varepsilon C \hat{q}_j^{\varepsilon-1}.
\end{equation*}
Also note that equation (\ref{eq:agg_cons}) together with the pricing equation (\ref{eq:price}) imply 
\begin{equation*}
    Q=\left(\int_{\mathcal{N}} p_j^{1-\varepsilon}dj\right)^{\frac{1}{1-\varepsilon}}=
    C^{\frac{1}{\varepsilon}}\left(\int_{\mathcal{N}}{c_j^{\frac{\varepsilon-1}{\varepsilon}}dj}\right)^{\frac{\varepsilon}{\varepsilon-1}}=C^{\frac{2}{\varepsilon}} = 1
\end{equation*}

Finally, plugging equation (\ref{eq:eqi_price_cons}) into equation (\ref{eq:agg_cons}) by using the definition of $Q$ in equation (\ref{eq:prod_index}) provides the wage for unskilled labor 
\begin{equation}
    w^u = \frac{\varepsilon-1}{\varepsilon}Q.
    \label{eq:wages_unsk}
\end{equation}
Note that here the dynamic of eventual endogenous exit becomes evident. As $Q$ increases over time due to innovation, so does unskilled wage, causing the relative productivity of a good $\hat{q}_j$ in equation (\ref{eq:rel_prod}) to decrease, given a constant productivity level $q_j$ (i.e. assuming no innovation in good $j$) up to a point where the production of that good is no longer profitable. To state this explicitly
\begin{equation*}
    \text{for}\:t,s\in\mathbb{R}^+:s>t\;\text{and}\;q_j(t)=q_j(s)\Rightarrow\lim_{s\rightarrow\infty}\hat{q}_j(s) = 0
\end{equation*} 
which assures eventual endogenous exit given positive fixed costs (i.e. $\phi w^s >0$).

\subsection{Firm Dynamics}

The model allows for several transitions between firm types. To recall, a firm can be \textit{active} or \textit{inactive} and can be of type \textit{basic}, \textit{applied low} and \textit{applied high}. 

Firms or goods can transition between \textit{active} and \textit{inactive} in the following ways.

Firstly, a firm can be shocked out of the economy due to some exogenous destructive unpredictable event. This happens at the exogenously derived rate $\delta$.

Secondly, following the literature of creative destruction (e.g. \cite{aghion1990model}, \cite{klette2004innovating})
a firm can be outperformed by a competitor at the endogenous rate $\tau$. If this occurs, it loses the produced product and exits the market.

Thirdly, a firm can endogenously exit the market, as over time the produced good loses relative profitability (recall $\frac{\partial\hat{q}(w^s)}{\partial w^s}\frac{\partial w^s(Q)}{\partial Q}\frac{\partial Q(t)}{\partial t}\leq 0$). Every good which is not innovated on will decay in profitability over time and eventually become unprofitable with probability one.

The forth transition between active and inactive firms is endogenous entrance, as described above using the decision tree in figure \ref{fig:innov_tree}. Entrants of whichever type need to hire $h^{e}=G(x^e,\theta^e)$ skilled labor as defined in equation (\ref{eq:rd_cost}) in order to innovate at rate $x^e$. Recall, selecting $\theta^e$ is stochastic as described in equation (\ref{eq:entry_prob}). Entrants thereby face the maximization problem
\begin{equation}
    \max_{x^{e}\geq0}{\left\{x^{e}\mathbb{E}V^{e}\left(\hat{q}(t+),\theta\right)-w^s G\left(x^{e},\theta^e\right)\right\}}.
    \label{eq:maxim_entrant}
\end{equation}
Here, $\mathbb{E}V^{e}\left(\hat{q}(t+),\theta\right)$ constitutes the expected value of entry and $w^s G\left(x^{e},\theta^e\right)$ is the cost of research when innovating at rate $x^e$ as described in equation (\ref{eq:rd_cost}) for $n=1$.\\

Further, the model allows for two transitions \textit{within} active firms, namely a transition from \textit{applied high} to \textit{applied low} and a transition from \textit{basic} to \textit{applied low}. The prior addresses the empirically observed tendency of firms losing innovative potential throughout their lifespan (\cite{aghion2015industrial},\cite{acemoglu2018innovation}), whereas the latter models the ability of basic research to become outdated (\cite{akcigit2021back}). These transitions happen at a flow rate $\nu$ and $\mu$ respectively and their mechanisms will become clear in the next section.

\subsection{Value Functions}
To determine the value for a firm of given type first note that each firm is maximizing its overall profits, i.e. profits of production and expected profits of research jointly. Further, the value function is bounded from below by zero which constitutes the outside option of being inactive. In order to find a stationary equilibrium all growing variables are normalized by the productivity index $Q(t)$ defined in equation (\ref{eq:prod_index}). Normalized variables are denoted by a tilde, e.g. profits $\pi(t)$ when normalized by $Q(t)$ are written as $\Tilde{\pi}(t)$. The value function for the absorbing type \textit{applied} \textit{low} R\&D takes the form 

\begin{equation}
    \begin{split}
        &r{\widetilde{V}}_{a,l}\left(\hat{\mathcal{Q}}\right)=\\
        &\max\Bigg\{0,\max_{x\geq0}\Bigg[\sum_{\hat{q}\in\hat{\mathcal{Q}}}\left[\widetilde{\pi}\left(\hat{q}\right)-{\widetilde{w}}^s\phi+\tau\left[{\widetilde{V}}_{a,l}\left(\hat{\mathcal{Q}}\backslash \left\{\hat{q}\right\}\right)-{\widetilde{V}}_{a,l}\left(\hat{\mathcal{Q}}\right)\right]+\frac{\partial{\widetilde{V}}_{a,l}\left(\hat{\mathcal{Q}}\right)}{\partial\hat{q}}\frac{\partial\hat{q}}{\partial w^u}\frac{\partial w^u}{\partial t}\right]\\
        &+n\left[x\left[\mathbb{E}{\widetilde{V}}_{a,l}\left(\hat{\mathcal{Q}}\cup\left\{\hat{q}+(\xi\eta +(1-\xi)\lambda)\bar{\hat{q}}\right\}\right) -{\widetilde{V}}_{a,l}\left(\hat{\mathcal{Q}}\right)\right]-{\widetilde{w}}^sG\left(x,\theta^{a,l}\right)\right]\\
        &+\varphi\left[-{\widetilde{V}}_{a,l}\left(\hat{\mathcal{Q}}\right)\right]\Bigg]\Bigg\}.
    \end{split}
    \label{eq:value_la}
\end{equation}

Starting from the inner maximization, a firm maximizes the sum of all the profits from the products it currently produces. Explicitly, given again some product line $j$, 
\begin{equation*}
    \widetilde{\pi}\left(\hat{q}\right)-{\widetilde{w}}^s\phi
\end{equation*} 
constitutes the profit on variable production minus fixed costs of management. 

The second expression 
\begin{equation*}
    \tau\left[{\widetilde{V}}_{a,l}\left(\hat{\mathcal{Q}}\backslash \left\{\hat{q}\right\}\right)-{\widetilde{V}}_{a,l}\left(\hat{\mathcal{Q}}\right)\right]
\end{equation*} 
represents the change in value when losing some product line due to creative destruction, which happens at a rate $\tau$, the rate of creative destruction. For clarity I would like to note here that $\tau=\bar{x}$ which is derived below, with $\bar{x}$ being the average rate of innovation in the economy, which aligns with intuition as successful innovation on an active product by one firm translates into creative destruction of another. 
The third expression in the sum,
\begin{equation*}
    \frac{\partial{\widetilde{V}}_{a,l}\left(\hat{\mathcal{Q}}\right)}{\partial\hat{q}}\frac{\partial\hat{q}}{\partial w^u}\frac{\partial w^u}{\partial t}
\end{equation*} 
reflects the change in a firm's value over time which is negative due to $\frac{\partial\hat{q}}{\partial w^u}<0$ by definition of $\hat{q}\equiv\frac{q}{w^s}$. 

In the second row of equation (\ref{eq:value_la}) the sum over all products can be replaced by $n$ as invention is identical for all products held. Similar to the entrance case in equation (\ref{eq:maxim_entrant}) $\mathbb{E}\Tilde{V}_{a,l}$ is the firm's expected gain in value in case a good of productivity $\left[\hat{q}+(\xi\eta +(1-\xi)\lambda)\bar{\hat{q}}\right]$ is added to the production portfolio, which happens at rate $x$. The last expression in the second row, 
\begin{equation*}
    \widetilde{w}^sG\left(x,\theta^{a,l}\right)
\end{equation*}
is structurally identical to (\ref{eq:maxim_entrant}) and constitutes the total cost of skilled labor employed to produce $x$. Finally,
\begin{equation*}
    \varphi\left[-{\widetilde{V}}_{a,l}\left(\hat{\mathcal{Q}}\right)\right]
\end{equation*}
models the exogenous destructive shock, which happens at rate $\varphi$ and comes at a cost of the current firm value.

Consequently, the value function of an \textit{applied high} type firm takes a similar structure, i.e
\begin{equation}
    \begin{split}
        &r{\widetilde{V}}_{a,h}\left(\hat{\mathcal{Q}}\right)=\\
        &\max\Bigg\{0,\max_{x\geq0}\Bigg[\sum_{\hat{q}\in\hat{\mathcal{Q}}}\left[\widetilde{\pi}\left(\hat{q}\right)-{\widetilde{w}}^s\phi+\tau\left[{\widetilde{V}}_{a,h}\left(\hat{\mathcal{Q}}\backslash\left\{\hat{q}\right\}\right)-{\widetilde{V}}_{a,h}\left(\hat{\mathcal{Q}}\right)\right]+\frac{\partial{\widetilde{V}}_{a,h}\left( \hat{\mathcal{Q}}\right)}{\partial\hat{q}}\frac{\partial\hat{q}}{\partial w^u}\frac{\partial w^u}{\partial t}\right]\\
        &+n\left[x\left[\mathbb{E}{\widetilde{V}}_{a,h}\left(\hat{\mathcal{Q}}\cup\left\{\hat{q}+(\xi\eta +(1-\xi)\lambda)\bar{\hat{q}}\right\}\right) -{\widetilde{V}}_{a,h}\left(\hat{\mathcal{Q}}\right)\right]-{\widetilde{w}}^sG\left(x,\theta^{a,h}\right)\right]\\
        &+\varphi\left[-{\widetilde{V}}_{a,h}\left(\hat{\mathcal{Q}}\right)\right]
        +\nu\left[{\widetilde{V}}_{a,l}\left(\hat{\mathcal{Q}}\right)-{\widetilde{V}}_{a,h}\left(\hat{\mathcal{Q}}\right)\right]\Bigg]\Bigg\}.
    \end{split}
    \label{eq:value_ha}
\end{equation}
In addition to the modeled dynamics in equation (\ref{eq:value_la}) the value of the transition from \textit{applied} \textit{high} to \textit{applied} \textit{low} type is expressed in the last term in the third row. The change in value of the firm
\begin{equation*}
    {\widetilde{V}}_{a,l}\left(\hat{\mathcal{Q}}\right)-{\widetilde{V}}_{a,h}\left(\hat{\mathcal{Q}}\right)
\end{equation*} 
at a given time happens with rate $\nu$.

The same dynamic applies to the basic research type. Note further, that the expected value of innovation is no longer dependent on active firms of either type as it is the case for the equations (\ref{eq:value_la}) and (\ref{eq:value_ha}). Innovation now happens at a fixed innovation step size $\eta$ as it is not effected by the spillovers of other research, may it be applied or other basic research. Explicitly, the value function takes the form 
\begin{equation}
    \begin{split}
        &r{\widetilde{V}}_b\left(\hat{\mathcal{Q}}\right)=\\
        &\max\Bigg\{0,\max_{x\geq0}\Bigg[\sum_{\hat{q}\in\hat{\mathcal{Q}}}\left[\widetilde{\pi}\left(\hat{q}\right)-{\widetilde{w}}^s\phi+\tau\left[{\widetilde{V}}_b\left(\hat{\mathcal{Q}}\backslash\left\{\hat{q}\right\}\right)-{\widetilde{V}}_b\left(\hat{\mathcal{Q}}\right)\right]+\frac{\partial{\widetilde{V}}_b\left(\hat{\mathcal{Q}}\right)}{\partial\hat{q}}\frac{\partial\hat{q}}{\partial w^u}\frac{\partial w^u}{\partial t}\right]\\
        &+n\left[x\left[\mathbb{E}{\widetilde{V}}_b\left(\hat{\mathcal{Q}}\cup\left\{\hat{q}+\eta\bar{\hat{q}}\right\}\right) -{\widetilde{V}}_b\left(\hat{\mathcal{Q}}\right)\right]-{\widetilde{w}}^sG\left(x,\theta^b\right)\right]\\
        &+\varphi\left[-{\widetilde{V}}_b\left(\hat{\mathcal{Q}}\right)\right]
        +\mu\left[{\widetilde{V}}_{a,l}\left(\hat{\mathcal{Q}}\right)-{\widetilde{V}}_b\left(\hat{\mathcal{Q}}\right)\right]\Bigg]\Bigg\}
    \end{split}
    \label{eq:value_b}
\end{equation}
where $\mu$ is the rate at which the basic research of the firm is outdated and it becomes a firm conducting \textit{low} \textit{applied} research.

It is important to realize that given the above value functions one can write the value of a firm as the sum of the values of each product line held by that firm, i.e.
\begin{equation*}
    {\widetilde{V}}_k\left(\hat{\mathcal{Q}}\right)=\sum_{\hat{q}\in\hat{\mathcal{Q}}}{\Upsilon^k\left(\hat{q}\right)}.
\end{equation*}
The detailed derivation for that finding can be found in the appendix. This property enriches the model substantially as firms can be arbitrarily broken down into sub-firms of heterogeneous nature. In the following the type of research employed depends on the good produced and thus each firm holds exactly one good which determines the type of research conducted. 

The value functions change slightly with the main structure unchanged. The value of a firm after having dropped a given product, which at an aggregated level used to be denoted by $\Tilde{V}_k(\hat{\mathcal{Q}}\backslash\{\hat{q}\})$ is now zero as it is the only product held. Similarly $\mathbb{E}{\widetilde{V}}_{a,l}\left(\hat{\mathcal{Q}}\cup\left\{\hat{q}(t+)\right\}\right)$ breaks down to $\mathbb{E}\Upsilon_{a,l}\left(\hat{q}(t+)\right)$. Further, $\max_{x\geq 0}$ can be rewritten so that the value function for e.g. the applied low type takes the form

\begin{equation}
    r\Upsilon_{a,l}\left(\hat{q}\right)=
    \begin{cases}
        \widetilde{\pi}\left(\hat{q}\right)-{\widetilde{w}}^s\phi+\tau\left[-\Upsilon_{a,l}\left(\hat{q}\right)\right]+\varphi\left[-\Upsilon_{a,l}\left(\hat{q}\right)\right]+\frac{\partial\Upsilon_{a,l}\left( \hat{q}\right)}{\partial\hat{q}}\frac{\partial\hat{q}}{\partial w^u}\frac{\partial w^u}{\partial t}+\Omega^{a,l}&\:\hat{q}>\hat{q}_{a,l,min}\\
        0 &\:\hat{q}\leq\hat{q}_{a,l,min}
    \end{cases}
    \label{eq:value_low_applied_to_be_sure}
\end{equation}

with 
\begin{equation}
    \Omega^{a,l}\equiv \max_{x\geq0}\Bigg[x\left[\mathbb{E}\Upsilon_{a,l}\left(\hat{q}+(\xi\eta +(1-\xi)\lambda)\bar{\hat{q}}\right) \right]-{\widetilde{w}}^sG\left(x,\theta^{a,l}\right)\Bigg]
    \label{eq:max_x_la}
\end{equation}
being the expected R\&D value of type $k\in\left\{\{a,l\},\{a,h\},\{b\}\right\}$.

Now recall that endogenous exit happens if $0\leq\hat{q}\leq \hat{q}_{min}$. Consequently, per-good value functions (henceforth simply called 'value functions') of active firms can be rewritten for the \textit{applied} \textit{low} type as 

\begin{equation}
    \begin{split}
        \left(r+\tau+\varphi\right)\Upsilon^{a,l}-\frac{\partial\Upsilon^{a,l}\left(\hat{q}\right)}{\partial\hat{q}}\frac{\partial\hat{q}}{\partial w^u}\frac{\partial w^u}{\partial t}=&\widetilde{\pi}\left(\hat{q}\right)-{\widetilde{w}}^S\phi+\Omega^{a,l}\\
        &\text{for}\quad \hat{q}>{\hat{q}}_{a,l,min},
    \end{split}
    \label{eq:value_fin_al}
\end{equation}
for the applied high type as
\begin{equation}
    \begin{split}
        \left(r+\tau+\varphi\right)\Upsilon^{a,h}-\frac{\partial\Upsilon^{a,h}\left(\hat{q}\right)}{\partial\hat{q}}\frac{\partial\hat{q}}{\partial w^u}\frac{\partial w^u}{\partial t}=&\widetilde{\pi}\left(\hat{q}\right)-{\widetilde{w}}^S\phi+\Omega^{a,h}+\nu\left[\Upsilon^{a,l}\left(\hat{q}\right)-\Upsilon^{a,h}\left(\hat{q}\right)\right]\\
        &\text{for}\quad \hat{q}>{\hat{q}}_{a,h,min},
    \end{split}
    \label{eq:value_fin_ah}
\end{equation}
and for the basic type as
\begin{equation}
    \begin{split}
        \left(r+\tau+\varphi\right)\Upsilon^b-\frac{\partial\Upsilon^b\left(\hat{q}\right)}{\partial\hat{q}}\frac{\partial\hat{q}}{\partial w^u}\frac{\partial w^u}{\partial t}=&\widetilde{\pi}\left(\hat{q}\right)-{\widetilde{w}}^S\phi+\Omega^b+\mu\left[\Upsilon^{a,l}\left(\hat{q}\right)-\Upsilon^b\left(\hat{q}\right)\right]\\
        &\text{for}\quad \hat{q}>{\hat{q}}_{b,min}.
    \end{split}
    \label{eq:value_fin_b}
\end{equation}

Writing equation (\ref{eq:max_x_la}) more generally for all types, where $k$ indicates the firm type and by using equation (\ref{eq:inov_st_with_spill}) provides
\begin{equation*}
    \Omega^k\equiv\max_{x\geq0}{\left\{x\mathbb{E}\Upsilon^k\left(\hat{q}(t+),\theta\right)-\Tilde{w}^sG\left(x,\theta^k\right)\right\}}.
\end{equation*}
When solving for the optimum $x$ for each type, we obtain
\begin{equation}
    x^k=\arg \max_{x\geq 0} \Omega^k=\theta^k\left[\frac{\left(1-\gamma\right)\mathbb{E}\Upsilon^k\left(\hat{q}+\lambda\bar{\hat{q}}\right)}{\Tilde{w}}^S\right]^{\frac{1-\gamma}{\gamma}}.
    \label{eq:eqi_innov_intens}
\end{equation}
Lastly, an analytical solution for the productivity threshold can be found. Knowing\\ $\Upsilon^k(\hat{q}_{k,min})=0$ allows to solve the value functions above and find for all three cases that
\begin{equation}
    {\hat{q}}_{k,min}=\left(\frac{{\widetilde{w}}^S\phi-\Omega^k}{\Pi}\right)^{\frac{1}{\varepsilon-1}}\quad\text{with}\quad \Pi\equiv\frac{1}{\varepsilon-1}\left(\frac{\varepsilon-1}{\varepsilon}\right)^\varepsilon,\quad k\in\left\{\{a,l\},\{a,h\},\{b\}\right\}.
    \label{eq:min_produc}
\end{equation}

The proof for this is analogous to \textcite{acemoglu2018innovation} in their two case model. For completeness, the full proof is found in the appendix below.

It is very important to notice the condition on the productivity level for each of the above mentioned value functions. In the case of \textit{applied low} research the firm exits if $\hat{q}\leq\hat{q}_{a,l,min}$, which does not hold true for the other two types. Transition patterns change if e.g. an \textit{applied} \textit{high} firm produces a product which drops below $\hat{q}_{a,l,min}$ and it holds that $\hat{q}_{a,h,min}<\hat{q}_{a,l,min}$ then the product line is still produced, however transition within firm-types is no longer possible. This can be seen when explicitly solving the differential equation. \textit{Applied} \textit{high} firms have an explicit value function of form

\begin{equation}
    \Upsilon^{a,l}(\hat{q}) =\max\left\{0,\frac{\Pi{\hat{q}}^{\varepsilon-1}}{\Psi+g\left(\varepsilon-1\right)}\left(1-\left(\frac{{\hat{q}}_{a,l,min}}{\hat{q}}\right)^{\frac{\Psi+g\left(\varepsilon-1\right)}{g}}\right)+\frac{\Omega^{a,l}-\Tilde{w}^s\phi}{\Psi}\left(1-\left(\frac{{\hat{q}}_{a,l,min}}{\hat{q}}\right)^{\frac{\Psi}{g}}\right)\right\}
    \label{eq:detailed_value_function_al}
\end{equation}
where the maximization accounts for the lower bound of zero.

For \textit{applied} \textit{high} as well as \textit{basic} firms, the explicit value function is a three-part-function. For $\hat{q}_{k,min}\leq\hat{q}\leq\hat{q}_{a,l,min}$ it is

\begin{equation}
        \Upsilon^{k}(\hat{q}) =\frac{\Pi{\hat{q}}^{\varepsilon-1}}{\Psi+\iota+g\left(\varepsilon-1\right)}\left(1-\left(\frac{{\hat{q}}_{k,min}}{\hat{q}}\right)^{\frac{\Psi+\iota+g\left(\varepsilon-1\right)}{g}}\right)+\frac{\Omega^{k}-\Tilde{w}^s\phi}{\Psi+\iota}\left(1-\left(\frac{{\hat{q}}_{k,min}}{\hat{q}}\right)^{\frac{\Psi+\iota}{g}}\right),
        \label{eq:detailed_value_function_ah_b_above}
\end{equation}

for $\hat{q}\geq\hat{q}_{a,l,min}$

\begin{equation}
\begin{split}
    &\Upsilon^{k}(\hat{q}) =\\ 
    &\left\{\small
    \begin{split}
        \frac{\Pi{\hat{q}}^{\varepsilon-1}}{\Psi+\iota+g\left(\varepsilon-1\right)}\left(1-\left(\frac{{\hat{q}}_{k,min}}{\hat{q}}\right)^{\frac{\Psi+\iota+g\left(\varepsilon-1\right)}{g}}\right)+\frac{\Omega^{k}-\Tilde{w}^s\phi}{\Psi+\iota}\left(1-\left(\frac{{\hat{q}}_{k,min}}{\hat{q}}\right)^{\frac{\Psi+\iota}{g}}\right)\\
        + \frac{\Pi{\hat{q}}^{\varepsilon-1}}{\Psi+g\left(\varepsilon-1\right)}\left(1-\left(\frac{{\hat{q}}_{a,l,min}}{\hat{q}}\right)^{\frac{\Psi+g\left(\varepsilon-1\right)}{g}}\right)+\frac{\Omega^{a,l}-\Tilde{w}^s\phi}{\Psi}\left(1-\left(\frac{{\hat{q}}_{a,l,min}}{\hat{q}}\right)^{\frac{\Psi}{g}}\right)\\
        - \frac{\Pi{\hat{q}}^{\varepsilon-1}}{\Psi+\iota+g\left(\varepsilon-1\right)}\left(1-\left(\frac{{\hat{q}}_{a,l,min}}{\hat{q}}\right)^{\frac{\Psi+\iota+g\left(\varepsilon-1\right)}{g}}\right)-\frac{\Omega^{a,l}-\Tilde{w}^s\phi}{\Psi+\iota}\left(1-\left(\frac{{\hat{q}}_{a,l,min}}{\hat{q}}\right)^{\frac{\Psi+\iota}{g}}\right)
    \end{split}
    \right\}
\end{split}
\label{eq:detailed_value_function_ah_b}
\end{equation}
and zero otherwise.

Note that here in abuse of notation $k\in\{\{a,h\},\{b\}\}$, excluding the otherwise included \textit{applied low} type. Further $\iota=\nu \text{I}(k=\{a,b\})+\mu \text{I}(k=\{b\})$ where $\text{I}(\cdot)$ is the indicator function and $\Psi=r+\tau+\varphi$. The first and the second row of equation (\ref{eq:detailed_value_function_ah_b}) are almost identical to equation (\ref{eq:detailed_value_function_ah_b_above}) and (\ref{eq:detailed_value_function_al}) respectively. The presence of the second line represents the value of transitioning to the \textit{applied} \textit{low} type and the third row being part of the boundary condition and intuitively down weights the value of performing as an \textit{applied} \textit{low} type to account for uncertainty of the transition.
    
\subsection{Productivity Distribution}

In order to find a stationary productivity distribution for the overall productivity levels in the economy the property
\begin{equation*}
    F_t(\hat{q}) = F_{t+\Delta t}(\hat{q}) = F_t(\hat{q}(1-g\Delta t))-\tau\Delta t[F_t(\hat{q})-F_t(\hat{q}-\zeta \bar{\hat{q}})]
\end{equation*}
must hold, stating that inflows and outflows due to creative destruction balance over a given time interval $\Delta t$. Rearranging and taking the limit as $\Delta t \rightarrow 0$ yields
\begin{equation}
    g\hat{q}f(\hat{q})=\tau[F(\hat{q})-F(\hat{q}-\zeta \bar{\hat{q}})]
    \label{eq:prod_distr_all}
\end{equation}
where $\zeta$ is the economy wide innovation step average
\begin{equation*}
    \zeta=\frac{(\Phi^{a,l}+\Phi^{a,h})(\xi\eta + (1-\xi)\lambda)+\Phi^b\eta}{\Phi^{a,l}+\Phi^{a,h}+\Phi^b}.
\end{equation*}

While having mentioned it already in equation (\ref{eq:inov_st_with_spill}), $\Phi^k$ is the active share of firm type $k$ in the market, s.t. $\Phi^{a,l}+\Phi^{a,h}+\Phi^b+\Phi^{np}=1$, where $\Phi^{np}$ is the inactive share of firms.

For each firm type the condition for the non-normalized productivity distribution $\Tilde{F}_k$ is
\begin{equation*}
    \begin{split}
        \Tilde{F}_{a,l,t}(\hat{q})= & \Tilde{F}_{a,l,t}(\hat{q}(1+g\Delta t)) - \Tilde{F}_{a,l,t}(\hat{q}_{a,l,min}(1+g\Delta t))\\
        & + \tau^{a,l}\Delta t[F_t(\hat{q}-\zeta\bar{\hat{q}})-\Tilde{F}_{a,l,t}(\hat{q})-F_t(\hat{q}_{a,l,min}-\zeta\bar{\hat{q}})]\\
        &-\left(\tau^{a,l} +\varphi\right)\Delta t\Tilde{F}_{a,l,t}(\hat{q})\\
        &+\nu\Delta t[\Tilde{F}_{a,h,t}(\hat{q})-\Tilde{F}_{a,h,t}(\hat{q}_{a,l,min})]
        +\mu\Delta t[\Tilde{F}_{b,t}(\hat{q})-\Tilde{F}_{b,t}(\hat{q}_{a,l,min})]
    \end{split}
\end{equation*}
for the \textit{applied} \textit{low} type,
\begin{equation*}
    \begin{split}
        \Tilde{F}_{a,h,t}(\hat{q})= &\Tilde{F}_{a,h,t}(\hat{q}(1+g\Delta t)) - \Tilde{F}_{a,h,t}(\hat{q}_{a,h,min}(1+g\Delta t))\\
        & + \tau^{a,h}\Delta t[F_t(\hat{q}-\zeta\bar{\hat{q}})-\Tilde{F}_{a,h,t}(\hat{q})-F_t(\hat{q}_{a,h,min}-\zeta\bar{\hat{q}})]\\
        &-\left(\tau^{a,h}+\varphi+\nu\right)\Delta t\Tilde{F}_{a,h,t}(\hat{q})
    \end{split}
\end{equation*}
for the \textit{applied} \textit{high} type and
\begin{equation*}
    \begin{split}
        \Tilde{F}_{b,t}(\hat{q})= & \Tilde{F}_{b,t}(\hat{q}(1+g\Delta t)) - \Tilde{F}_{b,t}(\hat{q}_{b,min}(1+g\Delta t))\\
        & + \tau^{b}\Delta t[F_t(\hat{q}-\zeta\bar{\hat{q}})-\Tilde{F}_{b,t}(\hat{q})-F_t(\hat{q}_{b,min}-\zeta\bar{\hat{q}})]\\
        &-\left(\tau^{b}+\varphi+\mu\right)\Delta t\Tilde{F}_{b,t}(\hat{q})
    \end{split}
\end{equation*}
for the \textit{basic} type. Note how the inflows associated with $\mu$ and $\nu$ in the \textit{applied} \textit{low} type are found as outflows in the other two. Again, rearranging and taking the limit as $\Delta t\rightarrow 0$ provides the conditions for stationary equilibrium

\begin{equation}
    \begin{split}
       g\hat{q}\Tilde{f}_{a,l}(\hat{q})=&g\hat{q}_{a,l,min}\Tilde{f}_{a,l}(\hat{q}_{a,l,min})+(\tau^{a,h}+\tau^{b}+\varphi)\Tilde{F}_{a,l}(\hat{q})\\
       &-\tau^{a,l}[F(\hat{q}-\zeta \bar{\hat{q}})-F(\hat{q}_{a,l,min}-\zeta \bar{\hat{q}})-\Tilde{F}_{a,l}(\hat{q})]\\ &-\nu[\Tilde{F}_{a,h}(\hat{q})-\Tilde{F}_{a,h}(\hat{q}_{a,l,min})]
       -\mu[\Tilde{F}_{b}(\hat{q})-\Tilde{F}_{b}(\hat{q}_{a,l,min})]
    \end{split}
    \label{eq:stat_prod_dist_al}
\end{equation}
for the \textit{applied} \textit{low} type,
\begin{equation}
    \begin{split}
       g\hat{q}\Tilde{f}_{a,h}(\hat{q})=&g\hat{q}_{a,h,min}\Tilde{f}_{a,h}(\hat{q}_{a,h,min})+(\tau^{a,l}+\tau^{b}+\varphi+\nu)\Tilde{F}_{a,h}(\hat{q})\\
       &-\tau^{a,h}[F(\hat{q}-\zeta \bar{\hat{q}})-F(\hat{q}_{a,h,min}-\zeta \bar{\hat{q}})-\Tilde{F}_{a,h}(\hat{q})]
    \end{split}
    \label{eq:stat_prod_dist_ah}
\end{equation}
for the \textit{applied} \textit{high} type, and
\begin{equation}
    \begin{split}
       g\hat{q}\Tilde{f}_{b}(\hat{q})=&g\hat{q}_{b,min}\Tilde{f}_{b}(\hat{q}_{b,min})+(\tau^{a,l}+\tau^{a,h}+\varphi+\mu)\Tilde{F}_{b}(\hat{q})\\
       &-\tau^{b}[F(\hat{q}-\zeta \bar{\hat{q}})-F(\hat{q}_{b,min}-\zeta \bar{\hat{q}})-\Tilde{F}_{b}(\hat{q})]
    \end{split}
    \label{eq:stat_prod_dist_b}
\end{equation}
for the \textit{basic} type where for all three above equations $\tau^k=\Phi^k x^k+P(\theta=\theta^k)x^e$ and $\Phi^k=\Tilde{F}_k(\infty)$, for $k\in\left\{\{a,l\},\{a,h\},\{b\}\right\}$. 

Note here that the above definition of $\tau$ also implies that the average innovation rate $\bar{x} = \sum_k \tau^k$ equals the rate of creative destruction $\tau$ in the economy as previously announced.

To conclude, the model solves analytically for a stationary productivity distribution among firm types. These results are necessary to determine the labor market equilibrium as it depends on the firm types active and their represented share in the economy.

\subsection{Labor Market}

Recall the production function (\ref{eq:prod_fun}) which can be rewritten in terms of labor. Using equation (\ref{eq:prod_is_cons}), the corollary $y_j=c_j$ and equation (\ref{eq:eqi_price_cons}), the unskilled labor market clears under the condition
\begin{equation}
    \int_{\mathcal{N}}l^*_j d_j= \int_{\mathcal{N}}y^*_j q_j^{-1}d_j=\int_{\mathcal{N}}c^*_j q_j^{-1}d_j=\left(\frac{\varepsilon-1}{\varepsilon}\right)^\varepsilon\left(w^u\right)^{-\varepsilon}C\int_{\mathcal{N}}q^{\varepsilon-1} d_j = 1
    \label{eq:unskilled_labor_clearing}
\end{equation}
which using equations (\ref{eq:prod_index}) and (\ref{eq:wages_unsk}) results in the observation that
\begin{equation*}
    C=Q \quad\Rightarrow\quad Y=C=Q.
\end{equation*} 
For skilled labor, note that entrants have to be explicitly considered as opposed to the unskilled case, their R\&D decision is to be distinguished from incumbents. Where incumbents for each type demand demand $\Phi^k[G(x^k,\theta^k)+\phi]$, entrants demand $G(x^e,\theta^e)$. The intuition follows directly from the definition of $G(\cdot,\cdot)$ and $\phi$ as they are defined as the demand of skilled labor for a given set of parameters $x$ and $\theta$. This labor demand is then weighted by the market share of that type $\Phi^k$. Jointly, the skilled labor clearing condition is
\begin{equation}
    G(x^e,\theta^e) + \sum_{k}\Phi^k[G(x^k,\theta^k)+\phi]=L^s
    \label{eq:skilled_labor_clearing}
\end{equation}
for again $k\in\{\{a,l\},\{a,h\},\{b\}\}$. Note that it is the skilled labor clearing condition which requires the stationary productivity distribution. It is this property in the model that accounts for redistribution of labor among firms.

\subsection{Economic Growth}

The growth rate of the economy can be obtained by rewriting and integrating equation (\ref{eq:prod_distr_all}) such that
\begin{equation*}
    \mathbb{E}(\hat{q})\equiv \frac{\tau}{g}\int_0^\infty[F(\hat{q})-F(\hat{q}-\zeta \bar{\hat{q}})]d\hat{q}.
\end{equation*}
By recognising that, as $\hat{q}>0$ it holds, $\mathbb{E}(\hat{q})=\int_0^\infty[1-F(\hat{q})]d\hat{q}$ and consequently one can rearrange to
\begin{equation*}
    \mathbb{E}(\hat{q})\equiv \frac{\frac{\tau}{g}}{1+\frac{\tau}{g}}\int_0^\infty[1-F(\hat{q}-\zeta \bar{\hat{q}})]d\hat{q}.
\end{equation*}
A change of variable substituting $x$ for $\hat{q}+\zeta\bar{\hat{q}}$ provides
\begin{equation*}
    \mathbb{E}(\hat{q})\equiv \frac{\frac{\tau}{g}}{1+\frac{\tau}{g}}\int_{-\zeta\bar{\hat{q}}}^\infty[1-F(x)]d\hat{q}=\frac{\tau}{g}\zeta\bar{\hat{q}}.
\end{equation*}
Now recognising that in equilibrium $\bar{\hat{q}}=\mathbb{E}(\hat{q})$ one finally obtains.
\begin{equation}
    g=\tau\zeta
    \label{eq:growth}
\end{equation}
This result follows somewhat intuitively as the economic equilibrium growth rate is proportional to the average innovation strep size of the economy scaled by the rate of creative destruction, which, as shown above, is also equal to the average innovation rate $\bar{x}$.

\subsection{Welfare}

To determine the welfare change for a given policy the utility of the representative household is evaluated at time $t=0$ and calculated as in equation (\ref{eq:utility}).
To solve the initial value problem, the expected value of the relative productivity at time zero $\hat{q}_0$ is normalized to one, thus implying according to equation (\ref{eq:prod_index}) that the initial productivity index $Q_0=\Phi^{\frac{1}{\varepsilon-1}}$ equals the initial consumption aggregate $C_0$. Thereby the explicit form of initial utility $U_0$ is
\begin{equation*}
    U_0(C_0,g) = \int_0^\infty e^{-\rho t}\frac{C_{t}^{1-\vartheta}-1}{1-\vartheta} dt = \frac{1}{1-\vartheta}\left[\frac{\Phi_0^{\frac{1-\vartheta}{\varepsilon-1}}}{\rho-g(1-\vartheta)}-\frac{1}{\rho}\right].
\end{equation*}
To compare two states of the economy both being in equilibrium, e.g. the market solution $s_0$ against a subsidy $s_1$, the consumption equivalent at time zero is computed given a changed growth rate $g(s_i)$, which is the fraction $\sigma$ of consumption that allows for the same utility given $g(s_i)$. This fraction $\sigma$ is found in the case of state $s_0$ and $s_1$ such that
\begin{equation*}
    U_0(\sigma C_0(s_1),g(s_1))=U_0(C_0(s_0),g(s_0))
\end{equation*}
holds.

\subsection{Stationary Equilibrium}

To conclude this section recall that a stationary equilibrium is found in the following way. Firms select their production $y^*_j$, monopoly price $p^*_j$ and labor demand $l^*_j$ according to (\ref{eq:eqi_price_cons}) and (\ref{eq:prod_fun}). The values for $\{\Upsilon^{a,l},\Upsilon^{a,h},\Upsilon^{b}\}$ derived from these choices are according to the value functions (\ref{eq:value_fin_al}), (\ref{eq:value_fin_ah}) and (\ref{eq:value_fin_b}). Equation (\ref{eq:min_produc}) provides the minimum productivities $\hat{q}_{k,min}$ for each type $k$ determining endogenous entry and exit. Given this market setup, all types select their innovation rate $x^k$ according to (\ref{eq:eqi_innov_intens}) and entrants solve their entry problem in (\ref{eq:maxim_entrant}). With a given innovation rate, skilled labor demand $h^k$ is selected using (\ref{eq:rd_cost}). To clear the skilled labor market (\ref{eq:stat_prod_dist_al}), (\ref{eq:stat_prod_dist_ah}) and (\ref{eq:stat_prod_dist_b}) are met to determine $\{\Phi^{a,l},\Phi^{a,h},\Phi^{b},\Phi^{np}\}$. Knowing each type's market share allows to evaluate the economy's growth rate $g$ by (\ref{eq:growth}). Wages $w^u$ and $w^s$ are set to meet the labor market clearing conditions in (\ref{eq:unskilled_labor_clearing}) and (\ref{eq:skilled_labor_clearing}) which affects the profits of firms. The interest rate $r$ is found to meet the Euler equation (\ref{eq:euler_eq}). This circular line of causation is used to solve for a fixed point.

\section{Computation of Equilibrium}\label{sec:Comp}

The model is solved for a fixed point on the discounted skilled labor wage rate $\Tilde{w}^s$, the active share of a firm type $\{\Phi^{a,l},\Phi^{a,h},\Phi^{b}\}$, the expected value after successful innovation $\{\mathbb{E}[\Upsilon^{a,l}(\hat{q}+(\xi\eta + (1-\xi)\lambda)\bar{\hat{q}})],\mathbb{E}[\Upsilon^{a,h}(\hat{q}+(\xi\eta + (1-\xi)\lambda)\bar{\hat{q}})],\mathbb{E}[\Upsilon^{b}(\hat{q}+\eta\bar{\hat{q}})]\}$, as well as the average relative productivity of the economy $\bar{\hat{q}}$. 
This constitutes the vector of endogenous equilibrium variables:
\begin{equation}
    \{\Tilde{w}^s,\Phi^{a,l},\Phi^{a,h},\Phi^{b},\mathbb{E}[\Upsilon^{a,l}(\hat{q}+(\xi\eta + (1-\xi)\lambda)\bar{\hat{q}})],\mathbb{E}[\Upsilon^{a,h}(\hat{q}+(\xi\eta + (1-\xi)\lambda)\bar{\hat{q}})],\mathbb{E}[\Upsilon^{b}(\hat{q}+\eta\bar{\hat{q}})],\bar{\hat{q}}\}.
    \label{eq:model_fp_variables}
\end{equation}

Using some initial guess for the above mentioned variables in equation (\ref{eq:model_fp_variables}) and the exogenous model primitives described in table \ref{tab:primitives}, the algorithm first maximises the expected value of R\&D $\{\Omega^{a,l},\Omega^{a,h},\Omega^{b}\}$ by finding the optimal innovation rate  $\{x^{a,l},x^{a,h},x^{b}\}$ which then allows to solve for the average creative destruction rate $\{\tau^{a,l},\tau^{a,h},\tau^{b}\}$, the minimum productivity threshold $\{\hat{q}_{a,l,min},$ $\hat{q}_{a,h,min},\hat{q}_{b,min}\}$ and aggregate growth rate $g$. Subsequently, the unnormalized productivity distributions $\{F,\Tilde{F}_{a,l},\Tilde{F}_{a,h},\Tilde{F}_{b}\}$ and the market shares $\{\Phi^{a,l},\Phi^{a,h},\Phi^{b},\Phi^{np}\}$ as well as the economy wide average productivity $\bar{\hat{q}}$ are obtained. Thereafter, the economy interest rate $r$ is determined which then allows to solve for the expected value functions $\{\mathbb{E}[\Upsilon^{a,l}],\mathbb{E}[\Upsilon^{a,h}],\mathbb{E}[\Upsilon^{b}]\}$. Finally the total demand of skilled labor $L^s$ is computed which concludes one iteration.

As the mapping from the complete variable space $\mathcal{V}(\cdot,\cdot)\rightarrow\mathcal{V}(\cdot,\cdot)$, spanned by the variables in (\ref{eq:model_fp_variables}) maps onto itself and constitutes a contraction mapping, conditional on the model primitives a unique fixed point exists (cf. \cite{banach1922operations}).

\begin{table}[]
\centering
\begin{tabular}{cllcc}
\hline
\hline\\ [-1.5ex]
Parameter & Description & Value & \multicolumn{2}{l}{Std.Err} \\
\\ [-1.5ex]
\hline
 $\alpha$      & Probability of being high type                         & 0.926 & (0.023) \\
 $\beta$       & Probability of being basic type                        & 0.100 & - \\
 $\phi$        & Cost of operation and management                       & 0.216 & (0.012) \\
 $\varphi$     & Exogenous rate of destruction                          & 0.037 & (0.001) \\
 $\nu$         & Transition rate from high- to low appl. type           & 0.206 & (0.005) \\
 $\mu$         & Transition rate from basic to low appl. type  (cool-down rate)         & 0.116 & *** \\
 $\lambda$     & Innovation step size for appl. without basic spillovers& 0.132 & (0.010) \\
 $\eta$        & Innovation step size for basic research                & 0.219 & *** \\
 $\theta^{a,l}$& Innovation capacity for appl. low type                 & 1.391 & (0.017) \\
 $\theta^{a,h}$& Innovation capacity for appl. high type                & 1.751 & (0.020) \\
 $\theta^{b}$  & Innovation capacity for basic type                     & 0.681 & *** \\
 $\theta^{e}$  & Innovation capacity for entrants                       & 0.024 & (0.001) \\
 \hline\\ [-1.5ex]
\end{tabular}
\caption{Parameterization taken from literature (model primitives)}
\label{tab:primitives}
\end{table}

The model primitives are taken from the literature. As the aforementioned papers\footnote{\textcite{acemoglu2018innovation} and \textcite{akcigit2021back}} base on most importantly \textcite{lentz2008empirical} parameters can translated with some confidence\footnote{Please note that this solution is recognized to be sub-optimal. The exogenous parameters would ideally be estimated by means of simulated methods of moments as was done by \textcite{akcigit2021back} and \textcite{acemoglu2018innovation} but due to the inaccessibility of micro-data this was not possible. The point estimates of the two models are estimated on different economies over a different time span. The majority of point estimates is taken from \textcite{acemoglu2018innovation}, where the estimated standard error is reported for completeness. The remaining parameters are taken from \textcite{akcigit2021back} and translated as explained. The behaviour of the standard error when translated is unknown as the SMM procedure only considers it's moment space, hence translated covariances cannot be estimated when translating into the \textcite{acemoglu2018innovation} environment.}. The only parameter which cannot be be translated from the academic literature is the probability of an entrant performing basic research ($\beta$). As of now, I am not aware of any indicator of this variable in the literature. The National Science Foundation (NSF) estimates the share basic research conducted in the US to be 17\% in 2021, 58\% of that being non-governmental resulting in approximately 10\% of non-governmental basic research. However, basic research is represented among entrants remains unclear. Without being able to estimate this, I assume the share of entrants conducting basic research to be 10\%.

The other three additional parameters are translated from \textcite{akcigit2021back}. First, the innovation step size ratio is taken to be constant. They find $\frac{\eta'}{\lambda'}=\frac{0.079}{0.049}=1.61$ and is taken to be in the same proportion in this model. Therefore, taking $\lambda=0.132$ provides $\eta = 1.61\lambda = 0.219$. This can be interpreted as innovation in basic research on average increases the productivity by $21.9\%$ as opposed to applied research which without further spillovers increase productivity by $13.2\%$ on average. Note that the applied research step size is estimated for all firms in the market in \textcite{acemoglu2018innovation} which tends to overestimates the innovation step size of applied firms. However, the exact numerical values are less important than the relation and going in line with intuition $\eta>\lambda$, where $\lambda$ is substantially smaller than $\eta$.

Secondly, the basic research cool-down rate $\mu$ is the flow rate from basic to applied type. Due to convergence reasons in the fixed point iteration, this model cannot allow for transition to \textit{applied} \textit{high} type but only to \textit{applied} \textit{low} type which does not violate intuition as there is no reason to believe why redundant basic research should continue to be highly innovative. Therefore the missing transition channel is not considered a shortcoming in the model. \textcite{akcigit2021back} provide an estimate for the cool-down rate in the economy which is exactly adopted here as its function in their paper is almost identical. It is 0.11 in value and thus approximately half of the transition rate from \textit{high} \textit{applied} to \textit{low} \textit{applied} type, i.e. basic research lasts longer in effect than highly productive applied research which is faster exhausted.

Thirdly, the innovation capacity of basic type firms can be translated directly from \textcite{akcigit2021back} as the innovation intensity is identically computed to \textcite{acemoglu2018innovation} and is found in equation (\ref{eq:inov_intens}). Note that \textcite{akcigit2021back} the R\&D cost function
is defined as
\begin{equation*}
    C_i(x,n,\Xi)=wnx_i^{\frac{1}{1-\gamma}}\Xi_i\quad\text{for}\quad i\in\{a,b\}
\end{equation*}
where $\{a,b\}$ corresponds to applied and basic research respectively, which implies $\Xi_i = (\theta^*_i)^{\frac{\gamma}{1-\gamma}}$ when translating it into the \textcite{acemoglu2018innovation} model. \textcite{akcigit2021back} find $\Xi_b=5.437$ thus implying $\theta^*_b=0.429$ and $\Xi_a=1.288$, yielding $\theta^*_a=0.902$. To fit these findings to the results of \textcite{acemoglu2018innovation} the relative productivity of basic research w.r.t. applied research is calculated as $\frac{\theta^*_b}{\theta^*_a}=0.475$. Now taking the estimated productivity average of active firms
\begin{equation*}
    \theta_a=\theta_h\frac{\Phi^h}{1-\Phi^{np}}+\theta_l\frac{\Phi^l}{1-\Phi^{np}}=1.432
\end{equation*}
results in $\theta_b=\frac{\theta^*_b}{\theta^*_a}\theta_a=0.681$ which is the value considered for this model lining up with intuition as the instantaneous capacity of innovation for basic research is smaller than for any applied firm, following the reasoning that uncertainty and implementation of basic innovation adds to the cost of basic R\&D.

\section{Results}\label{sec:Res}

In this section the model implied results for the market solution as well as possible policy treatments are discussed.

\subsection{Market Solution}

Using the above discussed parameter selection in table \ref{tab:primitives} the model solves for the parameter set in equation (\ref{eq:model_fp_variables}). Recall that the firms are selected into types with a given probability and their survival depends on the profitability of the good produced as well as their ability to innovate. The variables of interested obtained in the fixed point for equation can be obtained in table \ref{tab:market_sol} where all results for the sake of readability are represented in percentage points. Recall $x^k$ is the innovation intensity conducted by a firm type, $\Phi^k$ is the share of active firms in the economy of a given type, $\hat{q}_{k,min}$ constitutes the productivity threshold, $\frac{L^{R\&D}}{L^S}$ is the share of skilled labor employed in R\&D rather than management, $\tau=\bar{x}$ represents the rate of creative destruction and the average innovation intensity, $g$ is the growth rate and 'welf.' is the welfare which is normalized to one in the market solution (baseline) to allow for comparison in the subsequent analysis. 

\begin{table}[]
\centering
\begin{tabular}{cccccccccccccc}
\hline
\hline\\ [-1.5ex]
$x^e$ & $x^{a,l}$   & $x^{a,h}$  & $x^b$ & $\Phi^{a,l}$ & $\Phi^{a,h}$ & $\Phi^{b}$ & $\hat{q}_{a,l,min}$ & $\hat{q}_{a,h,min}$  & $\hat{q}_{b,min}$  & $\frac{L^{R\&D}}{L^S}$ & $\tau$ & g & welf. \\
\\ [-1.5ex]
\hline
\multicolumn{14}{l}{\textbf{Baseline:}}\\
[+1.5ex]
0.51 & 25.98 & 38.32 & 14.43 & 55.24 & 5.89 & 0.17 & 147.07 & 129.86 & 155.15 & 19.88 & 17.14 & 2.27 & 100.00 \\   
\hline\\ [-1.5ex]
\end{tabular}
\caption{Market Solution}
\label{tab:market_sol}
\end{table}

\begin{figure}
    \centering
    \includegraphics[scale = 0.6]{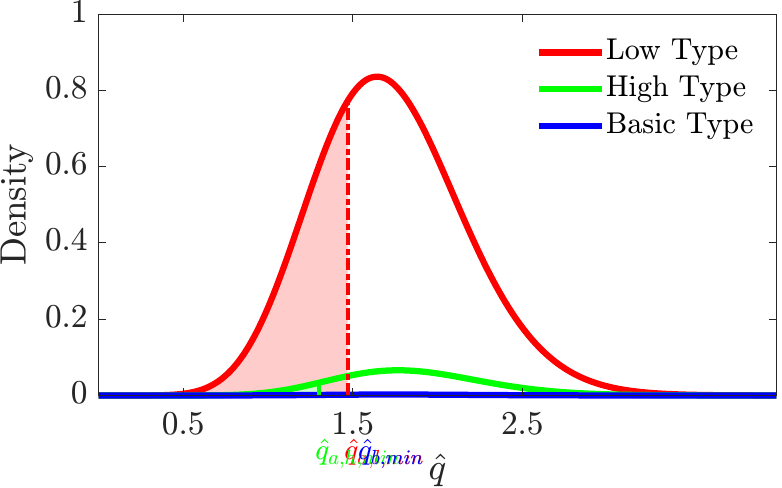}
    \includegraphics[scale = 0.6]{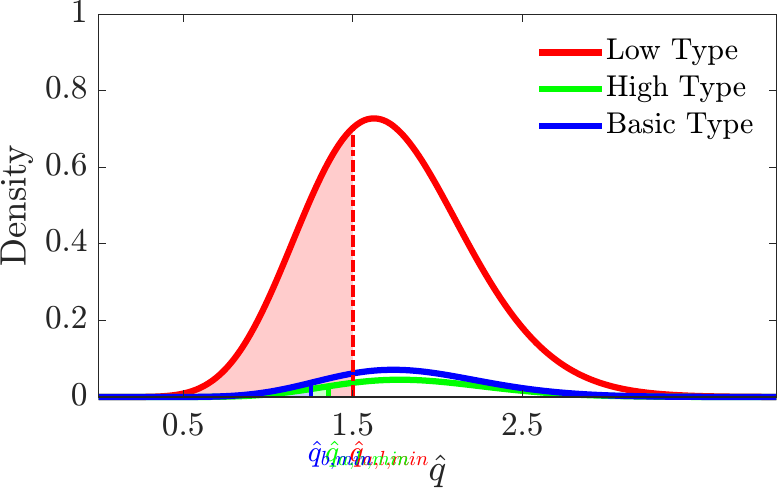}
    \caption{Stationary productivity distribution for the market solution (left) and a 1\% R\&D subsidy for incumbents (right)}
    \label{fig:market_sol}
\end{figure}

The market solution shows that high performing applied firms are on average $\frac{38.32}{25.98}=1.47$ more innovative than low applied ones and basic research is almost half as productive in innovation as low applied research\footnote{It is important to recall that innovation rate $x$ represents the flow rate of successful innovation for a given type and number of researchers employed and is not to be mistaken with the interpretation that basic research is less innovative with respect to quality.}. The share of research types in the economy is found to be mainly dominated by low applied research (90\% of active product lines are produced by that type). The remaining share is almost exclusively highly productive applied research (9.6\%). The basic research share in an open market economy finds its equilibrium value to be $0.28\%$ of active firms. The reason for this is that basic research is highly unprofitable compared to other types. It innovates at a higher cost and is not able to incorporate those cost due to creative destruction. Applied firms use the basic research profitability increase of that good and innovate on that, making basic research profits without baring the high R\&D costs. This environment also explains the larger relative profitability threshold $\hat{q}_{b,min}$, which implies firms conducting basic research leave the market earlier due to lack of profitability. Applied high research firms on the other hand are found to have a substantially lower threshold for endogenous exit compared to the applied low type. The threshold value of $\hat{q}_{a,h,min}$ is 11.7\% lower than $\hat{q}_{a,l,min}$. The share of skilled labor which is employed in R\&D rather than management is 19.88\%. Finally the rate of creative destruction $\tau$ takes the value of 17.14\% and equilibrium growth is 2.27\%. The left graph in figure \ref{fig:market_sol} visualizes the unnormalized stationary productivity distribution obtained by the system of delayed differential equations as expressed in (\ref{eq:stat_prod_dist_al}), (\ref{eq:stat_prod_dist_ah}) and (\ref{eq:stat_prod_dist_b}). The shaded area constitutes the unprofitable share of product lines associated with each research type.

Given a practically non-existent basic research sector, it should not be surprising that the results are similar to \textcite{acemoglu2018innovation} as the majority of primitives are taken from their paper. The solution also is in line with the economic intuition implying shortage of basic research in free markets as outlined in the introduction which boosts the confidence in the model structure.

\subsection{Research Subsidy for Incumbents}

Consider now a policy which addresses to subsidize the cost of research. The corresponding structure of R\&D cost is found in equation (\ref{eq:rd_cost}), and takes the form 
\begin{equation*}
    C(x,n,\theta) = w^s n x^{\frac{1}{1-\gamma}}\theta^{-\frac{\gamma}{1-\gamma}}.
\end{equation*}
The R\&D subsidy is proportional to the economy's GDP and increases the innovation capacity of firms. This follows the reasoning that by subsidizing a firm, research becomes relatively more productive in monetary terms which is modelled as 
\begin{equation*}
    \theta(1-s_{inc})^{-\frac{1-\gamma}{\gamma}}
\end{equation*}
where $(1-s_{inc})^{-\frac{1-\gamma}{\gamma}}$ weights the innovation capacity $\theta$ such that costs are equal to the determined subsidy size. The cost of the subsidy is
\begin{equation}
    C_{inc}=s_{inc}w^s\sum_k G(x^k,\theta^k)\Phi^k
    \label{eq:subs_inc_cost}
\end{equation}
so it is the cost of each type of research $k\in\{\{a,l\},\{a,h\},\{b\}\}$ that is happening due to the subsidy. These costs, together with the productivity increase, determine the net welfare change as described above. Note that if the subsidy is directed towards a specific type of research the running variable $k$ in equation (\ref{eq:subs_inc_cost}) changes from the full set of research types to the respective subset.\\

\subsubsection{Incumbent R\&D Subsidy for all Types}

Subsidizing each type with an equal share, i.e. having the same $s_{inc}$ for all types results in a new equilibrium summarized in table \ref{tab:rnd_subs_cases}, which shows that a R\&D subsidy benefits, as might be expected, high performing applied firms more than low applied types as they employ research more efficiently and hence have higher profits and lower exits. This comes at the cost of low applied firms as well as basic research firms who face a higher demand in skilled labor and thus a higher skilled wage rate making it more difficult for them to be profitable. This effect also reflects in the increase of productivity thresholds throughout all types. Higher demand in R\&D labor also explains the higher share of skilled labor employed in R\&D activities which, given a subsidy of 1\% of GDP, increases from 19.88\% in the market equilibrium to 21.78\%. The increase of the creative destruction rate is partly due to the fact that more research is done (which can be seen in the increase of all incumbent R\&D intensities $\{x^{a,l},x^{a,h},x^b\}$), implying more overtaking of active product lines by competitors as successful innovation happens more frequently. Since equilibrium growth is a function of R\&D intensity, it increases from 2.27\% to 2.35\%. Considering the cost of this measure calculated as in equation (\ref{eq:subs_inc_cost}), the overall increase in welfare is only 0.61\%. For entrants, the environment becomes less attractive. Higher demand for skilled labor increases its price, driving up entry costs and lowering the optimal innovation intensity $x^e$. Also note that subsidizing R\&D for incumbents and hence making entrance relatively more costly results in a lower share of total active firms which in the 1\% case decreases from 61.3\% to 59.85\%.\\


\subsubsection{Incumbent R\&D Subsidy for Applied Research}

Having evaluated a general subsidy without segregation between research types, a subsidy on applied research only has a very similar effect. This is little surprising as in the market solution hardly any basic research is applied. The results are found in table \ref{tab:rnd_subs_cases}.
Note first that both applied types have higher innovation intensity than in the overall subsidy which is unsurprisingly as the subsidy budget is now shared only between applied types which results in even lower R\&D costs for those types. This also explains the increased share of those types in the economy. The comparative decrease in $x^b$ and consequently $\Phi^b$ is explained by the fact that the increase in creative destruction which is largely unchanged to the unconditional subsidy is almost exclusively driven by applied types meaning basic types are outperformed and exit market more quickly. Also skilled wage increases on one side with no benefit for basic research on the other which further cuts the profitability of basic research showing in the increased $\hat{q}_{b,min}$. Welfare remains largely unchanged due to the overall little share of basic research, however missing the additional innovation with step size $\eta$ as well as spillovers from basic research attribute to a overall change in welfare by two basis points.\\

\begin{table}[]
\begin{tabular}{cccccccccccccc}
\hline
\hline\\ [-1.5ex]
$x^e$ & $x^{a,l}$   & $x^{a,h}$  & $x^b$ & $\Phi^{a,l}$ & $\Phi^{a,h}$ & $\Phi^{b}$ & $\hat{q}_{a,l,min}$ & $\hat{q}_{a,h,min}$  & $\hat{q}_{b,min}$  & $\frac{L^{R\&D}}{L^S}$ & $\tau$ & g & welf.\\
[-1.5ex]\\
\hline
\multicolumn{14}{l}{\textbf{Baseline:}}\\
[+1.5ex]
0.51 & 25.98 & 38.32 & 14.43 & 55.24 & 5.89 & 0.17 & 147.07 & 129.86 & 155.15 & 19.88 & 17.14 & 2.27 & 100.00 \\  
\hline
\multicolumn{14}{l}{\textbf{R\&D Incumbent Subsidy (1\% of GDP}):}\\
[+1.5ex]
\multicolumn{14}{l}{All Types:}\\
0.47 & 27.46 & 40.93 & 15.41 & 53.20 & 6.50 & 0.15 & 150.88 & 133.39 & 158.44 & 21.78 & 17.76 & 2.35 & 100.61\\
[+1.5ex]
\multicolumn{14}{l}{Applied Types Only:}\\
0.46 & 27.47 & 40.94 & 12.78 & 53.22 & 6.50 & 0.13 & 150.88 & 133.38 & 161.06 & 21.77 & 17.76 & 2.35 & 100.59\\
[+1.5ex]
\multicolumn{14}{l}{Basic Type Only:}\\
0.50 & 25.36 & 36.46 & 34.23 & 48.23 & 4.14 & 6.61 & 150.12 & 135.71 & 125.48 & 22.91 & 16.51 & 2.52 & 103.64\\
\hline \\ [-1.5ex]
\end{tabular}
\caption{Incumbent R\&D Subsidy.}
\label{tab:rnd_subs_cases}
\end{table}

\subsubsection{Incumbent R\&D Subsidy for Basic Research}

Subsidizing only basic research changes the picture substantially. The results of spending 1\% of GDP in such a policy are again stated in table \ref{tab:rnd_subs_cases}. First note that similar to the case of the applied R\&D subsidy, only the basic innovation intensity increases. This means that spillovers, as modelled above, are not strong enough to also increase innovation intensity for applied types, however, do partially counteract the decline in $x$.\footnote{To make a claim about the effects of a basic research subsidy, the structure of spillovers is essential. If e.g. the linear structure of spillovers is maintained but the weighting between $\lambda$ and $\eta$ is changed sufficiently adding more weight to $\eta$, a basic research subsidy is able to also increase the applied type innovation intensity, as will be shown below. Needless to say, the linear structure of spillovers in this model would require data to evaluate it's appropriateness and non-linear setups are equally possible and probably even more plausible, given potential interaction effects of multiple basic research achievements.} Increasing innovation capacity $\theta^b$ allows for an increase in basic research firms in the economy up to 6\% of total firms and 11.2\% of active firms. This comes at the cost of both applied low (-12.7\%) and applied high (-29.7
\%). The reason for low type firms being less effected is twofold. Firstly note that that now more basic research is happening which will eventually become outdated and transform to applied low types which feeds into the mass. At the same time the productivity thresholds rise for the applied types as skilled wage increases which hits highly productive applied types more as they employ more skilled labor to reach the desired innovation intensity $x$. This dynamic, in particular the effect of such a subsidy on productivity thresholds, depends on the parameterization of the model. With strong enough spillovers, the expected value of innovation is sufficiently enhanced so that it outweighs the rise in wage rate and thus also lowers the productivity thresholds for applied firms. The substantial innovation intensity ($x^b$) increase in basic research as well as the relatively small decrease in applied innovation intensity results in a high R\&D labor demand which is reflected in the increased R\&D labor ratio. To recall, this dynamic is determined by $G(x,\theta)$ in equation (\ref{eq:rd_cost}). The high productivity of basic research and the productivity increase via spillovers to applied research results in a growth rate of 2.52\% and a welfare gain of 3.64\%. Finally, the rate of creative destruction decreases which is due to a lower economy wide innovation rate. Again, this result depends on the parameterization but bears the intuition that higher skilled labor weges drives applied firms out of the market resulting in an overall total share of active goods or firms of 59.0\% as opposed to 61.3\% in the market solution. The average innovation intensity of active firms remains basically unchanged at 27.13\%, but the drop in the share of active firms results in an economy wide decrease (including inactive firms) of average innovation capacity i.e. creative destruction. In the right graph of figure \ref{fig:market_sol} the productivity distribution of such a measure is displayed. Note in particular the change of the ordering of productivity thresholds.

\subsection{Social Planner in a Market Solution}

In order to analyze the efficiency of the market equilibrium, a social planner is designed to calibrate the economy such that it provides a welfare maximizing solution. The social planner designs the market environment by changing the innovation intensities of each type as well as the endogenous exit thresholds, or explicitly the parameter set $\{x^{a,l},x^{a,h},x^b,\hat{q}_{a,l,min},\hat{q}_{a,h,min},\hat{q}_{b,min}\}$. This selection of parameters is done so that firms in their production process still behave optimally to the environment, i.e. no sub-optimal production decisions are made on a firm level. Thus, the social planner is to be considered similarly to an extremely strong state with a highly controlled firm environment with each firm still able to behave optimally.

The results of this optimization are listed in table \ref{tab:soc_plan_sol}. With the degrees of freedom the social planner has at her disposal, basic research cannot enter the market. Basic research remains at its innovation capacity $\theta^b$, regardless of what the social planner does. Shifting the endogenous exit rate $\hat{q}_{b,min}$ does not affect the incumbent's profitability and as entrants probabilities remain unchanged no extra inflow occurs. Basic research remains highly unattractive for firms and the increased demand in skilled labor by highly productive applied types increases skilled wage $w^s$ adding to this unprofitability. The social planner thus increases the productivity threshold and lowers the innovation rate of basic research firms to free skilled labor and redistribute it to applied high types which overall explains the decrease in the share of basic research firms in the social planner solution. For the same purpose, i.e. freeing skilled labor, low applied firms also face a (even) higher endogenous exit threshold but at the same time are required to have a slightly higher innovation capacity. This comes all to the benefit of highly productive applied research which is heavily encouraged to stay in the market by lowering $\hat{q}_{a,h,min}$ by approximately 81\%. The social planner solution results in a share of applied high research firms among all active firms of 87\% as opposed to 9.6\% in the market solution. At the same time market entry, which is not determined by the social planner, is increases to 0.58\% making it costly to enter the market given the increased skilled wage rate. Welfare increases by 4.95\% which shows the potential in redistributing skilled labor efficiently within applied types. The unnormalized stationary productivity distributions are displayed in the left graph of figure \ref{fig:soc_plan}.

To summarize, the results are very similar to \textcite{acemoglu2018innovation} as the social planner recognizes basic research to be unprofitable and consequently decides to set the innovation environment such that skilled labor is redistributed to more productive firm types. Spillovers are too week to make it profitable to increase the share of basic research and thus benefit applied types. Further, for a higher share in basic research to happen, skilled labor would be required. Facing inelastic labour supply, this would have to be taken from more productive types, harming welfare and resulting in a trap for basic research in the economy.

\begin{table}[]
\begin{tabular}{llllllllllllll}
\hline
\hline\\ [-1.5ex]
$x^e$ & $x^{a,l}$   & $x^{a,h}$  & $x^b$ & $\Phi^{a,l}$ & $\Phi^{a,h}$ & $\Phi^{b}$ & $\hat{q}_{a,l,min}$ & $\hat{q}_{a,h,min}$  & $\hat{q}_{b,min}$  & $\frac{L^{R\&D}}{L^S}$ & $\tau$ & g & welf.\\
[-1.5ex]\\
\hline
\multicolumn{14}{l}{\textbf{Baseline:}}\\
[+1.5ex]
\multicolumn{14}{l}{Market Equilibrium ($\theta^b=0.68$):}\\
0.51 & 25.98 & 38.32 & 14.43 & 55.24 & 5.89 & 0.17 & 147.07 & 129.86 & 155.15 & 19.88 & 17.14 & 2.27 & 100.00 \\
[+1.5ex]
\multicolumn{14}{l}{Equilibrium after R\&D Policy ($\theta^b=1.15$):}\\
0.52 & 25.86 & 37.91 & 32.31 & 53.66 & 5.53 & 1.62 & 147.62 & 131.10 & 126.69 & 20.53 & 17.02 & 2.33 & 100.99\\
\hline
\multicolumn{14}{l}{\textbf{Social Planner:}}\\
[+1.5ex]
\multicolumn{14}{l}{Market Equilibrium ($\theta^b=0.68$):}\\
0.58 & 27.54 & 45.43 & 9.94 & 6.62 & 43.67 & 0.13 & 234.81 & 46.59 & 176.49 & 34.13 & 22.25 & 3.04 & 104.59\\
[+1.5ex]
\multicolumn{14}{l}{Equilibrium after R\&D Policy ($\theta^b=1.15$):}\\
0.00 & 37.70 & 49.10 & 36.12 & 5.88 & 34.42 & 11.51 & 235.58 & 8.65 & 0.94 & 37.86 & 22.70 & 3.38 & 115.11\\
\hline \\ [-1.5ex]
\end{tabular}
\caption{Social Planner Solution}
\label{tab:soc_plan_sol}
\end{table}

\subsection{Social Planner Solution in a Subsidized Market}

With basic research being too unprofitable to be considered by a social planner in the market equilibrium, the question remains what an welfare optimal solution would be if a subsidy for basic research is already in place. This setup is to see whether and how the model accounts for crowding-in basic research and how a social planner would set the environment for monopolistic firms to optimize such an economy. In short, what is the potential of an economy where basic R\&D is subsidized compared to the optimal market solution and thus tackle the question of the necessity for governmental aid with respect to basic research. Therefore a subsidy of 0.25\% of GDP is considered which again addresses basic research only, to sufficiently crowd-in, so that spillovers can be taken advantage of and basic research is considered to be valuable for the social planner. The results of this optimization can be found in table \ref{tab:soc_plan_sol}. 

\begin{figure}
    \centering
    \includegraphics[scale = 0.6]{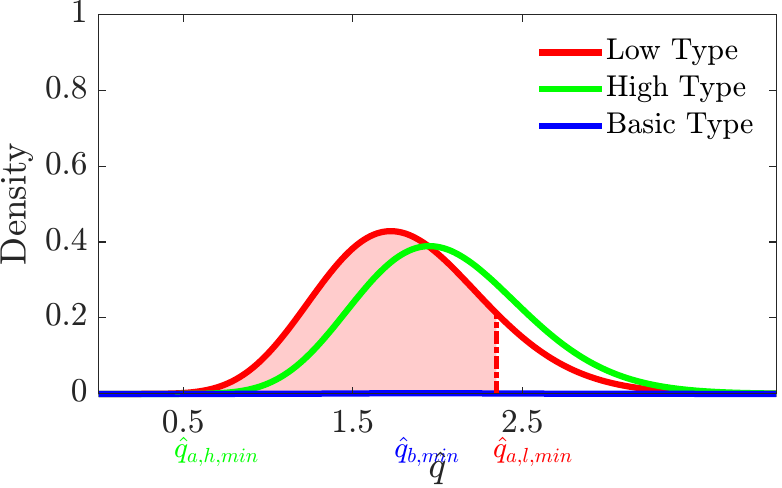}
    \includegraphics[scale = 0.6]{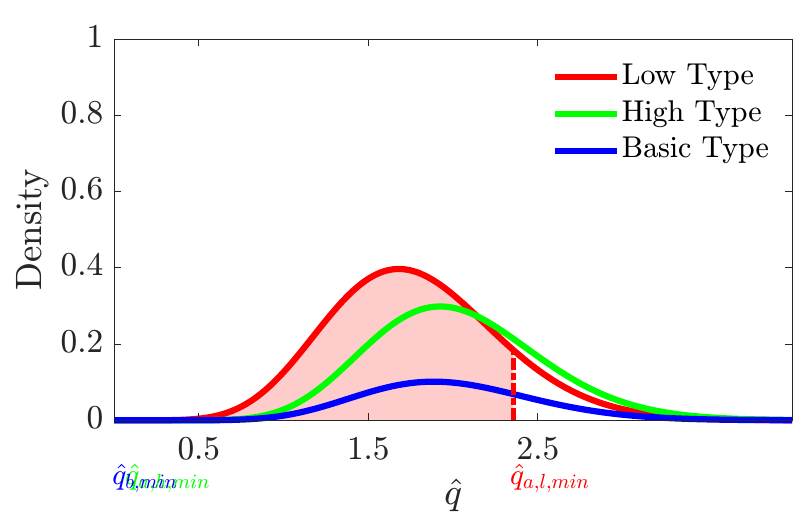}
    \caption{Social planner solution for the productivity distribution for the market case (left) and the equilibrium for a 0.2\% GDP basic research subsidy.}
    \label{fig:soc_plan}
\end{figure}

Given a basic R\&D policy in place, the welfare optimal innovation environment by the social planner is similar to the optimized market solution in case of the applied low type. This is little surprising as the underlying mechanisms are the same. However, with basic research sufficiently crowded-in by the subsidy, the share of active highly productive applied firms and basic research firms increases from approximately 44\% in the optimized market solution to 46\% in the subsidized market optimum. It is important to note there that by itself, basic research is less productive than the low applied type, and even is found to have a lower innovation intensity, but due to spillovers is encouraged to stay in the market for practically any given level of productivity. A similar environment is found for the high applied type, albeit less extreme. Further, all types are required to satisfy a relatively high rate of successful innovation, shifting up the R\&D labor ratio. The property of basic research to substantially increase productivity for each successful innovation ($\eta$) as well as its spillovers to applied types leads to an increase in creative destruction as well as a welfare gain of 15.11\%. Finally, entrants face a high skilled wage rate but at the same time extremely low minimum productivity thresholds for applied high and basic research. Given the entrant's probabilities of self-selecting into \textit{basic} and \textit{low} \textit{applied} are $\alpha$ and $\beta$ respectively, they only become low applied researchers in $(1-\alpha)(1-\beta)=6\%$ of the time. Therefore they innovate at a rate of almost zero, as $\hat{q}_{a,h,min},\hat{q}_{b,min}$ allow for almost all firms to remain in the market. The productivity distribution of the social planner equilibrium in a subsidized market is found in the right graph in figure \ref{fig:soc_plan}.  

To conclude, sufficient crowding-in of basic research increases the social optimum substantially, creating an economy mainly dominated by highly performing applied firms as well as basic research firms.

\section{Discussion}\label{sec:Dis}

The above results back the theoretical argument by e.g. \textcite{rosenberg2010firms} that basic research due to its characteristics is heavily under-supplied by the free market. Subsidizing research does not seem to solve this problem as long as it is not particularly tailored towards basic research as otherwise the comparative unprofitability remains. While a basic R\&D subsidy yields a substantial increase in welfare, the result falls short of the social planner. However, the social planner solution shows that it is most profitable for an economy for almost all basic research conducted to stay in the market regardless of their level of productivity. This result urges for solution beyond a free market, as interfering in the entry and exit behaviour to a required extent by an authority would entail substantial market distortions. A free lunch for entrants would be established as they face basically no entrance requirements. My work thus complements and supports \textcite{akcigit2021back} who model a governmental authority to bypass the issue of crowding-in. The authors setup however does not allow for the insight that sufficient crowding-in is necessary in order to take advantage of basic research and it's spillovers. My solution thus establishes the link between \textcite{acemoglu2018innovation} and \textcite{akcigit2021back} and provides a theoretical argument for the necessity of basic research conducted by the government.

Regarding the shortcomings of this model I would like to raise some points of concern. As frequently outlined above, the model calibration is done solely based on literature. This bears some apparent complications, as the parameters estimated in \textcite{acemoglu2018innovation} and \textcite{akcigit2021back} are estimated on different economies at different points in time. Where the prior focuses on US data in the period 1987-1997, the latter employs French data from 2000 to 2006, both on firm level. The employed method is simulated methods of moments (SMM) which is conducted by creating bins of the data set and estimate the first two centered moments jointly for all such data bins. The parameter estimation is then done such that the free parameters in table \ref{tab:primitives} are selected so that given the imposed model structure they fit the data moments best. As this implies knowing the covariance matrix of the data bins translating parameters, even if structurally correct, and assuming identical data sets is problematic as covariances are not taken into account. Having mentioned this, the full estimation of the model is possible as the required data is available for multiple countries.

Secondly, the modelling of the spillovers is arbitrary. The results show that spillovers have a substantial effect, in particular for the social planner solution after sufficient crowding-in by means of a basic R\&D subsidy has occurred. However, due to the lack of data the fit of model cannot be estimated and consequently the appropriateness of the spillover modelling remains unanswered. A possible extension of the spillover structure where the linear characteristic is maintained would be to introduce a weighting parameter $\varsigma\geq 1$ to change the spillovers to
\begin{equation}
    \xi = \frac{\varsigma\Phi^b}{\Phi^{a,l}+\Phi^{a,h}+\varsigma\Phi^{b}}
    \label{eq:new_spillover}
\end{equation}
where $\varsigma$ is the weighting parameter and would have to be estimated. In the baseline model described above $\varsigma=1$. The larger $\varsigma$, the heavier present basic research effects applied research in its innovation size. For example, if $\varsigma=20$\footnote{The number 20 is randomly chosen and has no further interpretation. It's mere purpose is to show there exists a $\varsigma>0$ such that a subsidy in basic research is beneficial to every firm type instantaneously.} the above claimed characteristic of spillovers to positively influence \textit{all} types of firms becomes apparent. The results are displayed in table \ref{tab:new_spill}.
Also, no distinction is made between spillovers to \textit{applied} \textit{high} and \textit{applied} \textit{low} research type respectively. Modelling such a property would allow for redistribution between applied researching types but would add additional parameters to the already heavily parameterized model which is why such an extension is not further examined. After all, the implications of such an amendment are foreseeable and it would have to be observable in the data that such an extension increases the model fit significantly.

\begin{table}[]
\begin{tabular}{llllllllllllll}
\hline
\hline\\ [-1.5ex]
$x^e$ & $x^{a,l}$   & $x^{a,h}$  & $x^b$ & $\Phi^{a,l}$ & $\Phi^{a,h}$ & $\Phi^{b}$ & $\hat{q}_{a,l,min}$ & $\hat{q}_{a,h,min}$  & $\hat{q}_{b,min}$  & $\frac{L^{R\&D}}{L^S}$ & $\tau$ & g & welf.\\
[-1.5ex]\\
\hline
\multicolumn{14}{l}{\textbf{Baseline:}}\\
[+1.5ex]
0.51 & 25.98 & 38.32 & 14.43 & 55.24 & 5.89 & 0.17 & 147.07 & 129.86 & 155.15 & 19.88 & 17.14 & 2.27 & 100.00 \\  
\hline
\multicolumn{14}{l}{\textbf{Heavier Spillovers ($\varsigma=20$):}}\\
[+1.5ex]
\multicolumn{14}{l}{Baseline:}\\
0.51 & 26.11 & 38.52 & 14.23 & 55.00 & 6.00 & 0.17 & 146.73 & 129.33 & 155.40 & 20.06 & 17.21 & 2.35 & 100.00\\
[+1.5ex]
\multicolumn{14}{l}{Equilibrium after 1\% Basic R\&D Policy:}\\
0.52 & 26.75 & 38.78 & 36.24 & 47.87 & 5.29 & 4.92 & 146.53 & 129.54 & 127.90 & 24.08 & 17.17 & 3.29 & 118.32\\
\hline \\ [-1.5ex]
\end{tabular}
\caption{Model implied results for stronger spillovers}
\label{tab:new_spill}
\end{table}

Thirdly, no transition between basic research and high performing applied research is modelled. The reason for this restriction lies in the structure of the value functions. As can be seen in equation (\ref{eq:detailed_value_function_ah_b}) a transition from e.g. applied high type to applied low type is only possible if the produced good of firm $f$ has a relative productivity higher than the endogenous exit threshold of the destination type, i.e. $\hat{q_f}\geq\hat{q}_{a,l,min}$. When modelling two transitions, e.g. \textit{basic} to \textit{high} \textit{applied} to \textit{low} \textit{applied}, the ordering of the respective productivity thresholds would have to be determined ex-ante, a restriction going against the nature of the model as the ordering of the endogenous exit thresholds are a key property.\footnote{One could design cases for all possible scenarios but this would destroy the convergence properties of the model as the value function changes throughout the value iteration.} However, this restriction is not necessarily essential. It implies that a firm conducting outdated basic research cannot become highly innovative applied research which makes intuitive sense as the innovation capacity of an exploited technology is low.

\section{Conclusion}\label{sec:Con}

I extended the model by \textcite{acemoglu2018innovation} by a basic research sector in order to evaluate the effect of a R\&D subsidy. Basic research is modeled such that it allows for a productivity spillovers from basic to applied research and is characteristically less innovative at each given point in time compared to applied research. The reason for this is the high level of uncertainty in conducting basic research, the inability to incorporate innovation expenses due to spillovers as well as the high costs after successful basic research to implement it in the production process. Due to the lack of data, the parameterization is taken from \textcite{acemoglu2018innovation} and the additional parameters are taken from \textcite{akcigit2021back} and translated into the employed model accordingly. 

The subsidy analyzed is such that a share of the costs of research for producing incumbent firms is covered at expense of the entire economy. The welfare effects of an undifferentiated policy is compared to one where only applied research is benefiting as well as to one which only addresses basic research. I find that given the aforementioned properties of basic research, the market solution for this environment barely includes any basic research conducted, as it is too unprofitable. The economy is heavily dominated by \textit{applied} \textit{low} types. When distinguishing between applied and basic types in the subsidy, I find that basic research subsidies have substantially more bite with respect to welfare. This effect is twofold. Firstly, more basic research increases growth due to the large innovation step size they take in case of successful innovation. Secondly, spillovers increase the innovation step size of applied research which makes R\&D more effective throughout the entire economy.

When analysing how a social planner would change the innovation environment, basic research still has no place. However, welfare can be substantially increased by setting high constrains on relatively \textit{applied} \textit{low} research, so that skilled labor is redistributed into the \textit{applied} \textit{high} type, which faces a comparatively lenient market environment.

This result begs the question what the social optimum would be if some basic research would be already in place, so that spillovers take effect. The model implied solution to a basic research subsidy of 0.2\% allows to sufficiently crowd-in basic research so that the social planner can take advantage of it's mutually beneficial properties. Even after controlling for the cost of such a subsidy, the social optimum is found to enhance welfare by roughly 15\% compared to the market social optimum without crowding-in. This effect is possible by freeing skilled labor in \textit{applied} \textit{low} applied firms and redistributing it in \textit{basic} and \textit{applied} \textit{high} firms. Due to the difficult environment faced by the \textit{applied} \textit{low} firms, their exit is encouraged. As basic and high applied firms transform into low applied research, it can also be seen an economy where old ideas in the form of innovation are quickly abandoned freeing resources for frontier research and its applied implementation.

To conclude, my results go in line with the theoretical argument suggested by the economic literature. An economy does not provide sufficient basic research if let alone in a free market. Even a social planner designing the innovation environment and endogenous exit and entry levels cannot change that. However, once basic research is sufficiently crowded-in, the welfare gains are substantial. It is left for further research to calibrate this model on a coherent data set. It is then that the exact spillover modelling can be determined and numerical interpretation of policy effects gain interpretability in size. Overall the model yields intuitive results and provides evidence for the necessity of a government to sufficiently crowd-in basic research so that spillover effects can be taken advantage of. Such a model structure can also be used for e.g. sustainable research and innovation management which seem to have a similar characteristic, due to the tendency of free markets to undersupply those. 

\printbibliography
\newpage


\appendix
\setcounter{equation}{0}
\numberwithin{equation}{section}
\textbf{Derive the explicit value functions:} To derive (\ref{eq:detailed_value_function_al}), (\ref{eq:detailed_value_function_ah_b_above}) and (\ref{eq:detailed_value_function_ah_b}) first note $\Tilde{\pi}=\left(\frac{\varepsilon-1}{\varepsilon}\right)^\varepsilon \frac{1}{\varepsilon-1}\hat{q}^{\varepsilon-1} = \Pi \hat{q}^{\varepsilon-1}$
and $\Psi\equiv r+\tau+\varphi$. Then equation (\ref{eq:value_low_applied_to_be_sure}) can be rewritten as
\begin{equation*}
    \Psi \Upsilon^{a,l}(\hat{q})+g\hat{q}\frac{\partial \Upsilon^{a,l}(\hat{q})}{\partial\hat{q}}=\Pi \hat{q}^{\varepsilon-1}-\Tilde{w}^s\phi + \Omega^{a,l} \qquad \text{if} \hat{q}>\hat{q}_{a,l,min}
\end{equation*}
which can be rewritten as
\begin{equation*}
    \zeta_1\hat{q}^{-1}\Upsilon^{a,l}(\hat{q})+\frac{\partial \Upsilon^{a,l}(\hat{q})}{\partial\hat{q}}= \zeta_2\hat{q}^{\varepsilon-2}-\zeta_3\hat{q}^{-1} 
\end{equation*}
where $\zeta_1\equiv\frac{\Psi}{g}$,$\zeta_2\equiv\frac{\Pi}{g}$ and $\zeta_3\equiv \frac{\Tilde{w}^s\phi + \Omega^{a,l}}{g}$. Solving the differential equation above yields
\begin{equation*}
    \Upsilon^{a,l}(\hat{q}) = \hat{q}^{-\zeta_1}\left(\int\left[\zeta_2 t^{\zeta_1+\varepsilon-2}-\zeta_3 t^{\zeta_1-1}\right]dt+D\right) = \frac{\zeta_2\hat{q}^{\varepsilon-1}}{\zeta_1+\varepsilon-1}-\frac{\zeta_3}{\zeta_1}+D\hat{q}^{-\zeta_1},
\end{equation*}
where $D$ is the constant of integration which can be obtained by solving for $\Upsilon^{a,l}(\hat{q}_{a,l,min})=0$. Doing so yields
\begin{equation*}
    D=-\frac{\zeta_2\hat{q}_{a,l,min}^{\zeta_1+\varepsilon-1}}{\zeta_1+\varepsilon-1}+\frac{\zeta_3\hat{q}_{a,l,min}^{\zeta_1}}{\zeta_1}.
\end{equation*}
Plugging into the above value function and rearranging provides 
\begin{equation*}
    \Upsilon^{a,l} =\frac{\Pi{\hat{q}}^{\varepsilon-1}}{\Psi+g\left(\varepsilon-1\right)}\left(1-\left(\frac{{\hat{q}}_{a,l,min}}{\hat{q}}\right)^{\frac{\Psi+g\left(\varepsilon-1\right)}{g}}\right)+\frac{\Omega^{a,l}-\Tilde{w}^s\phi}{\Psi}\left(1-\left(\frac{{\hat{q}}_{a,l,min}}{\hat{q}}\right)^{\frac{\Psi}{g}}\right)
\end{equation*}.

To derive the high applied type and the basic type value function it is helpful to rewrite the low applied value function as
\begin{equation*}
    \Upsilon^{a,l}(\hat{q})=\zeta_4\hat{q}^{\varepsilon-1}+\zeta_5\hat{q}^{-\frac{\Psi}{g}}-\zeta_6,
\end{equation*}
with 
\begin{equation}
    \zeta_4\equiv\frac{\Pi}{\Psi+g\left(\varepsilon-1\right)},\zeta_5\equiv \frac{\left(\Tilde{w}^s\phi-\Omega^{a,l}\right)\hat{q}_{a,l,min}^{\frac{\Psi}{g}}}{\Psi} - \frac{\Pi \hat{q}_{a,l,min}^{\frac{\Psi}{g}+\varepsilon-1}}{\Psi+g(\varepsilon-1)} \;\text{and}\; \zeta_6\equiv\frac{\Tilde{w}^s\phi-\Omega^{a,l}}{\Psi}.
    \label{eq:app_zeta_4_6}
\end{equation}\\

The applied high type as well as the basic type have a value function of the same structure. For simplicity, in abuse of notation $k\in\{\{a,h\},\{b\}\}$ and $\iota=\nu \text{I}(k=\{a,b\})+\mu \text{I}(k=\{b\})$ with $\text{I}(\cdot)$ being the indicator function. The value functions for both types take the form

\begin{equation*}
\begin{split}
    (\Psi+\iota)\Upsilon^{k}(\hat{q}) + g\hat{q}\frac{\partial \Upsilon^{k}(\hat{q})}{\partial\hat{q}}=&\Pi \hat{q}^{\varepsilon-1}-\Tilde{w}^s\phi + \Omega^{k}+\iota\left(\zeta_4 \hat{q}^{\varepsilon-1}+\zeta_5\hat{q}^{-\frac{\Psi}{g}}-\zeta_6\right)\quad \text{for}\quad \hat{q}\geq\hat{q}_{a,l,min}\\ 
    (\Psi+\iota)\Upsilon^{k}(\hat{q}) + g\hat{q}\frac{\partial \Upsilon^{k}(\hat{q})}{\partial\hat{q}}=&\Pi \hat{q}^{\varepsilon-1}-\Tilde{w}^s\phi + \Omega^{k}\quad \text{for}\quad \hat{q}_{a,l,min}\geq\hat{q}\geq\hat{q}_{k,min}
\end{split}
\end{equation*}

which can be written as
\begin{equation*}
    K_1\Upsilon^k(\hat{q})\hat{q}^{-1}+\frac{\partial\Upsilon^k(\hat{q})}{\partial\hat{q}}=K_2\hat{q}^{\varepsilon-2}+K_3\hat{q}^{-\frac{\Psi+g}{g}}-K_4  \hat{q}^{-1}
\end{equation*}
where depending on the case:\\
\begin{equation}
    K_1 \equiv \frac{\Psi+\iota}{g}, K_2 \equiv \frac{\Pi+\iota\zeta_4}{g}, K_3 \equiv \frac{\iota\zeta_5}{g} \;\text{and}\; K_4 \equiv \frac{\iota\zeta_6+\Tilde{w}^s\phi-\Omega^k}{g} \;\text{for}\; \hat{q}\geq\hat{q}_{a,l,min}
    \label{eq:app_Ks_larger}
\end{equation}\\
\begin{equation}
    K_1 \equiv \frac{\Psi+\iota}{g}, K_2 \equiv \frac{\Pi}{g}, K_3 \equiv 0 \;\text{and}\; K_4 \equiv \frac{\Tilde{w}^s\phi-\Omega^k}{g}\;\text{for}\; \hat{q}_{a,l,min}\hat{q}\geq\hat{q}_{k,min}
    \label{eq:app_Ks_between}
\end{equation}

The solution of this differential equation is then found to be
\begin{equation}
\begin{split}
    \Upsilon^k(\hat{q})=&\hat{q}^{-K_1}\left(\int\left[K_2\hat{q}^{K_1+\varepsilon-2}+K_3\hat{q}^{K_1-\frac{\Psi+g}{g}}-K_4\hat{q}^{K_1-1}\right]d\hat{q}+D\right)\\
    =& \frac{K_2\hat{q}^{\varepsilon-1}}{K_1+\varepsilon-1}+\frac{K_3\hat{q}^{1-\frac{\Psi+g}{g}}}{K_1+1-\frac{\Psi+g}{g}}-\frac{K_4}{K_1}+D\hat{q}^{-K_1}.
\end{split}
\label{eq:app_val_ah_b}
\end{equation}
Again finding the constant of integration by using $\Upsilon^k(\hat{q}_{k,min})=0$ yields
\begin{equation*}
    D=-\frac{K_2\hat{q}_{k,min}^{K_1+\varepsilon-1}}{K_1+\varepsilon-1}+\frac{K_3\hat{q}_{k,min}^{K_1+1-\frac{\Psi+g}{g}}}{K_1+1-\frac{\Psi+g}{g}} + \frac{K_4\hat{q}_{k,min}^{K_1}}{K_1}\quad\text{for}\quad\hat{q}_{k,min}\leq\hat{q}\leq\hat{q}_{l,min}
\end{equation*}

Plugging $D$ into (\ref{eq:app_val_ah_b}) using (\ref{eq:app_Ks_between}), rearranging provides the value function for both $k\in\{\{a,h\},\{b\}\}$ types which takes the form
\begin{equation*}
        \Upsilon^{k}(\hat{q}) =\frac{\Pi{\hat{q}}^{\varepsilon-1}}{\Psi+\iota+g\left(\varepsilon-1\right)}\left(1-\left(\frac{{\hat{q}}_{k,min}}{\hat{q}}\right)^{\frac{\Psi+\iota+g\left(\varepsilon-1\right)}{g}}\right)+\frac{\Omega^{k}-\Tilde{w}^s\phi}{\Psi+\iota}\left(1-\left(\frac{{\hat{q}}_{k,min}}{\hat{q}}\right)^{\frac{\Psi+\iota}{g}}\right)
\end{equation*}
for $\hat{q}_{k,min}\leq\hat{q}\leq\hat{q}_{l,min}$.\\

In the case of $\hat{q}\geq\hat{q}_{a,l,min}$ one obtains
\begin{equation*}
        \Upsilon^{k}(\hat{q}) =\frac{\Pi{\hat{q}}^{\varepsilon-1}}{\Psi+g\left(\varepsilon-1\right)}\left(1-\left(\frac{{\hat{q}}_{a,l,min}}{\hat{q}}\right)^{\frac{\Psi+g\left(\varepsilon-1\right)}{g}}\right)+\frac{\Omega^{a,l}-\Tilde{w}^s\phi}{\Psi}\left(1-\left(\frac{{\hat{q}}_{a,l,min}}{\hat{q}}\right)^{\frac{\Psi}{g}}\right)+\frac{\Omega^{k}-\Omega^{a,l}}{\Psi+\iota}+D\hat{q}^{-\frac{\Psi+\iota}{g}}
\end{equation*}
as well as the boundary condition
\begin{equation}
        \Upsilon^{k}(\hat{q}_{a,l,min}) =\frac{\Pi{\hat{q}_{a,l,min}}^{\varepsilon-1}}{\Psi+\iota+g\left(\varepsilon-1\right)}\left(1-\left(\frac{{\hat{q}}_{k,min}}{\hat{q}_{a,l,min}}\right)^{\frac{\Psi+\iota+g\left(\varepsilon-1\right)}{g}}\right)+\frac{\Omega^{k}-\Tilde{w}^s\phi}{\Psi+\iota}\left(1-\left(\frac{{\hat{q}}_{k,min}}{\hat{q}_{a,l,min}}\right)^{\frac{\Psi+\iota}{g}}\right).
        \label{eq:app_bound_larger}
\end{equation}
Using (\ref{eq:app_Ks_larger}) and (\ref{eq:app_val_ah_b}), the value function $\Upsilon^k$ at the point $\hat{q}_{a,l,min}$ for $\hat{q}\geq\hat{q}_{a,l,min}$ is

\begin{equation}
\begin{split}
    \Upsilon^k(\hat{q}_{a,l,min})=& \frac{K_2\hat{q}_{a,l,min}^{\varepsilon-1}}{K_1+\varepsilon-1}+\frac{K_3\hat{q}_{a,l,min}^{1-\frac{\Psi+g}{g}}}{K_1+1-\frac{\Psi+g}{g}}-\frac{K_4}{K_1}+D\hat{q}_{a,l,min}^{-K_1}.
\end{split}
\label{eq:app_val_ah_b_above}
\end{equation}

Now, (\ref{eq:app_val_ah_b_above}) has to equal (\ref{eq:app_bound_larger}). Solving for the constant of integration one obtains

\begin{equation}
    D=\left\{ 
    \begin{split}
        -\frac{\Pi}{\Psi+\iota+g(\varepsilon-1)}\hat{q}_{k,min}^{\frac{\Psi+\iota+g(\varepsilon-1)}{g}}+\frac{\Tilde{w}^s\phi-\Omega^k}{\Psi+\iota}\hat{q}_{k,min}^{\frac{\Psi+\iota}{g}}\\
        -\frac{\iota \zeta_4}{\Psi+\iota+g(\varepsilon-1)}\hat{q}_{a,l,min}^{\frac{\Psi+\iota+g(\varepsilon-1)}{g}}-\zeta_5\hat{q}_{a,l,min}^{\frac{\iota}{g}}+\frac{\iota \zeta_6}{\Psi+\iota}\hat{q}_{a,l,min}^{\frac{\Psi+\iota}{g}}
    \end{split}
    \right\}
\end{equation}
whereby after plugging into (\ref{eq:app_val_ah_b_above}) and by use of (\ref{eq:app_zeta_4_6}) and rearranging the explicit value function for $\hat{q}\geq\hat{q}_{a,l,min}$ is

\begin{equation}
\begin{split}
    &\Upsilon^{k} =\\ 
    &\left\{\small
    \begin{split}
        \frac{\Pi{\hat{q}}^{\varepsilon-1}}{\Psi+\iota+g\left(\varepsilon-1\right)}\left(1-\left(\frac{{\hat{q}}_{k,min}}{\hat{q}}\right)^{\frac{\Psi+\iota+g\left(\varepsilon-1\right)}{g}}\right)+\frac{\Omega^{k}-w^s\phi}{\Psi+\iota}\left(1-\left(\frac{{\hat{q}}_{k,min}}{\hat{q}}\right)^{\frac{\Psi+\iota}{g}}\right)\\
        + \frac{\Pi{\hat{q}}^{\varepsilon-1}}{\Psi+g\left(\varepsilon-1\right)}\left(1-\left(\frac{{\hat{q}}_{a,l,min}}{\hat{q}}\right)^{\frac{\Psi+g\left(\varepsilon-1\right)}{g}}\right)+\frac{\Omega^{a,l}-w^s\phi}{\Psi}\left(1-\left(\frac{{\hat{q}}_{a,l,min}}{\hat{q}}\right)^{\frac{\Psi}{g}}\right)\\
        - \frac{\Pi{\hat{q}}^{\varepsilon-1}}{\Psi+\iota+g\left(\varepsilon-1\right)}\left(1-\left(\frac{{\hat{q}}_{a,l,min}}{\hat{q}}\right)^{\frac{\Psi+\iota+g\left(\varepsilon-1\right)}{g}}\right)-\frac{\Omega^{a,l}-w^s\phi}{\Psi+\iota}\left(1-\left(\frac{{\hat{q}}_{a,l,min}}{\hat{q}}\right)^{\frac{\Psi+\iota}{g}}\right)
    \end{split}
    \right\}.
\end{split}
\label{eq:app_detailed_value_function_ah_b}
\end{equation}

The endogenous exit thresholds are now found by establishing the first order condition. For the low applied type this yields
\begin{equation*}
    \frac{\partial\Upsilon(\hat{q})}{\partial t }\biggr\rvert_{\hat{q}=\hat{q}_{a,l,min}}=\frac{1}{g}\left(\Pi\hat{q}_{a,l,min}^{\varepsilon-2}+\frac{\Omega^{a,l}-\Tilde{w}^s}{\hat{q}_{a,l,min}}\right)=0 \Rightarrow \hat{q}_{a,l,min}=\left(\frac{\Omega^{a,l}-\Tilde{w}^s}{\Pi}\right)^{\frac{1}{\varepsilon-1}}
\end{equation*}
and similarly for both high applied and basic type 
\begin{equation*}
    \frac{\partial\Upsilon(\hat{q})}{\partial t }\biggr\rvert_{\hat{q}=\hat{q}_{k,min}}=\frac{1}{g}\left(\Pi\hat{q}_{k,min}^{\varepsilon-2}+\frac{\Omega^{k}-\Tilde{w}^s}{\hat{q}_{k,min}}\right)=0 \Rightarrow \hat{q}_{k,min}=\left(\frac{\Omega^{k}-\Tilde{w}^s}{\Pi}\right)^{\frac{1}{\varepsilon-1}}
\end{equation*}
\begin{flushright} $\qed$ \end{flushright}
\newpage

\textbf{Proof} of ${\widetilde{V}}_k\left(\hat{\mathcal{Q}}\right)=\sum_{\hat{q}\in\hat{\mathcal{Q}}}{\Upsilon^k\left(\hat{q}\right)}$:\\

Note that by the definition of $\hat{\mathcal{Q}}$ it holds that
\begin{equation}
\begin{split}
    r{\widetilde{V}}_{a,l}\left(\hat{\mathcal{Q}}\right)=
        &\sum_{\hat{q}\in\hat{\mathcal{Q}}}\max\Bigg\{0,\max_{x\geq0}\Bigg[\widetilde{\pi}\left(\hat{q}\right)-{\widetilde{w}}^s\phi+\tau\left[-\Upsilon_{a,l}\left(\hat{q}\right)\right]+\varphi\left[-\Upsilon_{a,l}\left(\hat{q}\right)\right]\\
        &+\frac{\partial\Upsilon_{a,l}\left( \hat{q}\right)}{\partial\hat{q}}\frac{\partial\hat{q}}{\partial w^u}\frac{\partial w^u}{\partial t} +x\left[\mathbb{E}\Upsilon_{a,l}\left(\hat{q}+(\xi\eta +(1-\xi)\lambda)\bar{\hat{q}}\right) \right]-{\widetilde{w}}^sG\left(x,\theta^{a,l}\right)\Bigg]\Bigg\}.
\end{split}
\end{equation}
By acknowledging that firms may chose not to produce a certain product line directly implies by definition of $\Upsilon^{a,l}(\hat{q})$
\begin{equation}
\begin{split}
        r\Upsilon^{a,l}(\hat{q}) = &\max\Bigg\{0,\widetilde{\pi}\left(\hat{q}\right)-{\widetilde{w}}^s\phi+\tau\left[-\Upsilon_{a,l}\left(\hat{q}\right)\right]+\varphi\left[-\Upsilon_{a,l}\left(\hat{q}\right)\right]\\
        &+\frac{\partial\Upsilon_{a,l}\left( \hat{q}\right)}{\partial\hat{q}}\frac{\partial\hat{q}}{\partial w^u}\frac{\partial w^u}{\partial t} +\max_{x\geq0}\Bigg[x\left[\mathbb{E}\Upsilon_{a,l}\left(\hat{q}+(\xi\eta +(1-\xi)\lambda)\bar{\hat{q}}\right) \right]-{\widetilde{w}}^sG\left(x,\theta^{a,l}\right)\Bigg]\Bigg\}
\end{split}
\end{equation}
and thus it holds that
\begin{equation*}
    {\widetilde{V}}_k\left(\hat{\mathcal{Q}}\right)=\sum_{\hat{q}\in\hat{\mathcal{Q}}}{\Upsilon^k\left(\hat{q}\right)}
\end{equation*}
which concludes the proof.

The same reasoning trivially holds true for $\Upsilon^{a,h}(\hat{q})$ and $\Upsilon^{b}(\hat{q})$  when using the definitions in (\ref{eq:value_ha}) and (\ref{eq:value_b}).
\begin{flushright} $\qed$ \end{flushright}

\end{document}